\documentclass[12pt]{article}
\usepackage{graphicx}

\def\hybrid{\topmargin 0pt      \oddsidemargin 0pt
        \headheight 0pt \headsep 0pt
       \voffset-1cm
        \textwidth 6.25in       
       \textheight 9.5in       
        \marginparwidth 0.0in
        \parskip 5pt plus 1pt   \jot = 1.5ex}
\catcode`\@=11
\def\marginnote#1{}

\newcount\hour
\newcount\minute
\newtoks\amorpm
\hour=\time\divide\hour by60
\minute=\time{\multiply\hour by60 \global\advance\minute by-\hour}
\edef\standardtime{{\ifnum\hour<12 \global\amorpm={am}%
        \else\global\amorpm={pm}\advance\hour by-12 \fi
        \ifnum\hour=0 \hour=12 \fi
        \number\hour:\ifnum\minute<10 0\fi\number\minute\the\amorpm}}
\edef\militarytime{\number\hour:\ifnum\minute<10 0\fi\number\minute}

\def\draftlabel#1{{\@bsphack\if@filesw {\let\thepage\relax
   \xdef\@gtempa{\write\@auxout{\string
      \newlabel{#1}{{\@currentlabel}{\thepage}}}}}\@gtempa
   \if@nobreak \ifvmode\nobreak\fi\fi\fi\@esphack}
        \gdef\@eqnlabel{#1}}
\def\@eqnlabel{}
\def\@vacuum{}
\def\draftmarginnote#1{\marginpar{\raggedright\scriptsize\tt#1}}

\def\draftlabel#1{{\@bsphack\if@filesw {\let\thepage\relax
   \xdef\@gtempa{\write\@auxout{\string
      \newlabel{#1}{{\@currentlabel}{\thepage}}}}}\@gtempa
   \if@nobreak \ifvmode\nobreak\fi\fi\fi\@esphack}
        \gdef\@eqnlabel{#1}}
\def\@eqnlabel{}
\def\@vacuum{}
\def\draftmarginnote#1{\marginpar{\raggedright\scriptsize\tt#1}}

\def\draft{\oddsidemargin -.5truein
        \def\@oddfoot{\sl preliminary draft \hfil
        \rm\thepage\hfil\sl\today\quad\militarytime}
        \let\@evenfoot\@oddfoot \overfullrule 3pt
        \let\label=\draftlabel
        \let\marginnote=\draftmarginnote
   \def\@eqnnum{(\theequation)\rlap{\kern\marginparsep\tt\@eqnlabel}%
\global\let\@eqnlabel\@vacuum}  }

\def\numberbysection{\@addtoreset{equation}{section}
        \def\theequation{\thesection.\arabic{equation}}}

\def\underline#1{\relax\ifmmode\@@underline#1\else
        $\@@underline{\hbox{#1}}$\relax\fi}

\def\titlepage{\@restonecolfalse\if@twocolumn\@restonecoltrue\onecolumn
     \else \newpage \fi \thispagestyle{empty}\c@page\z@
        \def\thefootnote{\fnsymbol{footnote}} }

\def\endtitlepage{\if@restonecol\twocolumn \else  \fi
        \def\thefootnote{\arabic{footnote}}
        \setcounter{footnote}{0}}  
\relax

\numberbysection
\hybrid

\newfont{\Bbb}{msbm10 scaled 1\@ptsize00}
\newfont{\Bbbb}{msbm7 scaled 1\@ptsize00}
\newcommand{\CC}{\mbox{\Bbb C}}

\newcommand{\DDD}{\raise-1pt\hbox{$\mbox{\Bbbb D}$}}

\newcommand{\UUU}{\raise-1pt\hbox{$\mbox{\Bbbb U}$}}

\newcommand{\ZZ}{\mbox{\Bbb Z}}
\newcommand{\z}{\raise-1pt\hbox{$\mbox{\Bbbb Z}$}}

\newcommand{\sss}{\raise-1pt\hbox{$\mbox{\Bbbb S}$}}

\def\beq{\begin{equation}}
\def\eeq{\end{equation}}
\def\p{\partial}

\newtheorem{lemma-definition}{Lemma-Definition}[section]

\begin{document}

\begin{titlepage}

\title{Dispersionless version 
of multi-component \\Pfaff-Toda hierarchy}

\author{A. Savchenko\thanks{
Skolkovo Institute of Science and Technology, 143026, Moscow, Russia
and National Research University Higher School of Economics,
20 Myasnitskaya Ulitsa,
Moscow 101000, Russia,
e-mail: ksanoobshhh@gmail.com}
\and
A.~Zabrodin\thanks{
National Research University Higher School of Economics,
20 Myasnitskaya Ulitsa,
Moscow 101000, Russia and
NRC ``Kurchatov institute'', Moscow, Russia;
e-mail: zabrodin@itep.ru}}

\date{December 2025}
\maketitle

\vspace{-7cm} \centerline{ \hfill ITEP-TH-37/25}\vspace{7cm}

\begin{abstract}

We consider the dispersionless limit of the recently introduced multi-component
Pfaff-Toda hierarchy. Its dispersionless version is a set of nonlinear differential equations for the dispersionless limit of logarithm of the tau-function 
(the $F$-function). They are obtained as limiting cases of bilinear
equations of the Hirota-Miwa type. The analysis of the 
Pfaff-Toda hierarchy is substantially simplified by using the observation 
that the full (not only dispersionless)
$N$-component Pfaff-Toda hierarchy is actually equivalent to 
the $2N$-component DKP hierarchy. In the dispersionless limit, 
there is an elliptic curve built in the structure of the hierarchy, 
with the elliptic modular parameter being a dynamical variable.
This curve can be uniformized by elliptic functions, and in the elliptic parametrization the hierarchy acquires a compact and especially nice form.

\end{abstract}

\end{titlepage}

\vspace{5mm}

%

\tableofcontents

\vspace{5mm}

\section{Introduction}

Along with the well-known hierarchies of integrable 
Kadomtsev-Petviashvili (KP) and 2D Toda type equations, 
there are more general (and still less studied) Pfaffian versions of them, 
so named because some of their exact solutions are expressed not in terms of determinants, but in terms of Pfaffians. 
One such hierarchy, first briefly mentioned in 
the work of Jimbo and Miwa in 1983 \cite{JM83}, 
was rediscovered several times later and appears in the literature 
under more than one name (DKP, coupled KP, etc.), see \cite{HO}--\cite{Kodama}. 
The Pfaff-Toda hierarchy was introduced by Takasaki in \cite{Takasaki09}.
Recently, in our papers \cite{SZ24,SZ25a} multi-component generalizations
of these hierarchies were introduced. (For multi-component generalizations
of the KP and Toda hierarchies see \cite{DJKM81}--\cite{TZ25}.)

In the $N$-component Pfaff-Toda hierarchy, the independent variables are
the $N$ infinite sets of ``times'',
\beq\label{int1}
\begin{array}{l}
{\bf t}=\{{\bf t}_1, {\bf t}_2, \ldots , {\bf t}_N\}, \qquad
{\bf t}_{\alpha}=\{t_{\alpha , 1}, t_{\alpha , 2}, t_{\alpha , 3}, 
\ldots \, \},
\\ \\
\bar {\bf t}=\{\bar {\bf t}_1, \bar {\bf t}_2, \ldots , 
\bar {\bf t}_N\}, \qquad
\bar {\bf t}_{\alpha}=\{\bar t_{\alpha , 1}, \bar t_{\alpha , 2}, 
\bar t_{\alpha , 3}, \ldots \, \},
\end{array}
\qquad \alpha = 1, \ldots , N
\eeq
and two finite sets of discrete variables
$$
{\bf n}=\{n_1, \ldots , n_N\}, \quad
\bar {\bf n}=\{\bar n_1, \ldots , \bar n_N\}, \quad n_{\alpha}, \bar n_{\alpha}
\in \ZZ .
$$
In the bilinear formalism, the universal
dependent variable is the tau-function
$\tau ({\bf n}, \bar {\bf n}, {\bf t},
\bar {\bf t})$. The tau-function satisfies an infinite
number of bilinear equations which can be encoded in a single 
integral bilinear functional relation obtained
in \cite{SZ25a} (equation (\ref{main}) below in section
\ref{section:integral}). 

As is clear from the basic 
equation (\ref{main}), the $N$-component DKP hierarchy can be regarded as a part
of the Pfaff-Toda hierarchy 
obtained from it by freezing all bar-variables to zero values. On the
other hand, as we show in the present paper, the $N$-component Pfaff-Toda
hierarchy is equivalent to the $2N$-component DKP hierarchy, with the 
following renaming of the variables:
\beq\label{int2}
\left \{
\begin{array}{lll}
t^{\rm DKP}_{\alpha}=t^{\rm Toda}_{\alpha}, \quad
t^{\rm DKP}_{N+\alpha}=\bar t^{\rm Toda}_{\alpha},&&
\\ 
&& \quad \alpha =1, \ldots , N.
\\ 
n^{\rm DKP}_{\alpha}=n^{\rm Toda}_{\alpha}, \quad
n^{\rm DKP}_{N+\alpha}=\bar n^{\rm Toda}_{\alpha}&&
\end{array} \right.
\eeq
The corresponding tau-functions differ only by a sign factor (depending
on the discrete variables).
This resembles the correspondence between the multi-component KP and
Toda hierarchies \cite{UT84,TZ25}.

In this paper we are mostly interested in the dispersionless limit
of the mul\-ti-com\-po\-nent Pfaff-type hierarchies.
Following the approach developed in \cite{TT95},
we introduce an extra
parameter $\hbar \to 0$ 
and re-scale the times ${\bf t}$, $\bar {\bf t}$
and the discrete variables ${\bf n}$, $\bar {\bf n}$ as
$t_{\alpha , k}\to t_{\alpha , k}/\hbar$, 
$n_{\alpha}\to t_{\alpha , 0}/\hbar$ (and similarly 
for the bar-variables). Introduce the function 
$F({\bf t}_0, \bar {\bf t}_0,  {\bf t}, \bar {\bf t}; 
\hbar )$ related to the tau-function of the multi-component
Pfaff-Toda hierarchy by the
formula
\beq\label{d1a}
\tau \Bigl (\hbar^{-1}{\bf t}_0 , \hbar^{-1}\bar {\bf t}_0 , 
\hbar^{-1}{\bf t} , \hbar^{-1}\bar {\bf t} \Bigr )=
\exp \left ( \frac{1}{\hbar^2}\,
F({\bf t}_0, \bar {\bf t}_0, {\bf t},\bar {\bf t}; \hbar )\right )
\eeq
and consider the limit  
$$
F=F({\bf t}_0, \bar {\bf t}_0, {\bf t},\bar {\bf t})=
\lim\limits_{\hbar \to 0}
F({\bf t}_0, \bar {\bf t}_0, {\bf t},\bar {\bf t}; \hbar ).$$ 
The function $F$
plays the role of the 
tau-function in the dispersionless limit $\hbar \to 0$, meaning that
it serves as a universal dependent variable. 
The $F$-function satisfies 
an infinite number of highly nonlinear differential equations which are
limiting cases of the bilinear equations for the 
tau-function rewritten in terms of the
$F$-function. 

For the Pfaff-type hierarchies the dispersionless limit is especially 
interesting because the most important object in 
it is a certain elliptic curve built in the structure of the  hierarchy.
In the one-component case this phenomenon was noticed by Takasaki
\cite{Takasaki09,Takasaki07}. Later, in \cite{AZ14,AZ15} it was suggested
to uniformize the curve using elliptic functions. It was shown that
the uniformuzation allows one
to give a compact and nice formulation of the hierarchies. 
(A similar formulation exists for the usual dispersionless KP and 
Toda hierarchies, but there the elliptic curve 
degenerates to a rational one, which is uniformized by trigonometric functions,
see \cite{Z24}). 
In \cite{SZ24} this approach was applied to the dispersionless
multi-component DKP hierarchy, and it was shown that the plenty of
equations of the hierarchy
convert into a single one which has the following suggestive form:
\beq\label{int4}
\epsilon_{\beta \alpha}(a^{-1}-b^{-1})^{\delta_{\alpha \beta}} 
e^{\nabla_{\alpha}(a)\nabla_{\beta}(b)F}
= \frac{\theta_1(u_{\alpha}(a)- u_{\beta}(b)|\tau )}{\theta_4(u_{\alpha}(a)-
u_{\beta}(b)|\tau )} 
\eeq
for all $a,b \in \CC$ and
$\alpha , \beta =1, \ldots , N$.
In this form, the DKP (i.e. Pfaff-KP) hierarchy 
looks like an ``elliptic deformation'' of the KP one.
In (\ref{int4}), $\theta_1(u |\tau ), \theta_4(u|\tau )$ are the Jacobi
theta-functions with a modular parameter $\tau$,
\beq\label{n1c}
\nabla_{\alpha}(z)= \p_{t_{\alpha , 0}} +\sum_{k\geq 1}\frac{z^{-k}}{k}\,
\p_{t_{\alpha , k}}
\eeq
and $\epsilon_{\alpha \beta}$ is a sign factor equal to $1$ if 
$\alpha \leq \beta$ and $-1$ otherwise. For more details on the notation
see section \ref{section:elliptic}. The modular parameter $\tau =\tau ({\bf t})$
is a dynamical variable, and the dependence on times can be determined
from the equation
\beq\label{E11b}
e^{2\p_{\alpha}\p_{\beta}F}+e^{-2\p_{\alpha}\p_{\beta}F}-
e^{-2\p_{\alpha}^2F}(\p_{\beta}\p_{t_{\alpha ,1}}F)^2 =
\frac{\theta_2^2 (0|\tau )}{\theta_3^2 (0|\tau )}+
\frac{\theta_3^2 (0|\tau )}{\theta_2^2 (0|\tau )},
\eeq
where $\p_{\alpha}=\p_{t_{\alpha ,0}}$.

In section \ref{section:elliptic} we revisit the elliptic formulation
of the multi-component DKP hierarchy and suggest an approach which is
more general and simpler than the one used in \cite{SZ24}. The equivalence
of the DKP and Pfaff-Toda multi-component hierarchies mentioned above
allows us to obtain the result for the Pfaff-Toda case 
just by renaming the variables according to (\ref{int2}):
\beq\label{final1}
\left \{
\begin{array}{rcl}
\epsilon_{\beta \alpha}(a^{-1}-b^{-1})^{\delta_{\alpha \beta}} 
e^{\nabla_{\alpha}(a)\nabla_{\beta}(b)F}
&=& \displaystyle{\frac{\theta_1(u_{\alpha}(a)- 
u_{\beta}(b)|\tau )}{\theta_4(u_{\alpha}(a)-
u_{\beta}(b)|\tau )},}
\\ && \\
e^{\nabla_{\alpha}(a)\bar \nabla_{\beta}(b)F}& =& \displaystyle{
\frac{\theta_1(u_{\alpha}(a)+ \bar u_{\beta}(b)|\tau )}{\theta_4(u_{\alpha}(a)+
\bar u_{\beta}(b)|\tau )},}
\\ && \\
\epsilon_{\beta \alpha}(a^{-1}-b^{-1})^{\delta_{\alpha \beta}} 
e^{\bar \nabla_{\alpha}(a)\bar \nabla_{\beta}(b)F}
& =& \displaystyle{\frac{\theta_1(\bar u_{\alpha}(a)- 
\bar u_{\beta}(b)|\tau )}{\theta_4(\bar u_{\alpha}(a)-
\bar u_{\beta}(b)|\tau )}.}
\end{array} \right.
\eeq
For the one-component case it agrees with the result obtained
in \cite{AZ15}.

The structure of the paper is as follows. In section 2 we start by presenting the 
general bilinear integral equation for the tau-function obtained in \cite{SZ25a}
and establish the equivalence of the $N$-component Pfaff-Toda and 
$2N$-component DKP hierarchies. We also suggest a general method to 
obtain bilinear equations of the Hirota-Miwa type, and present some
details on the simplest meaningful case of 4-point relations.
Section 3 is devoted to the dispersionless limit, and the elliptic 
parametrization is discussed in detail. Section 4 contains concluding
remarks and some problems for further study. In the appendix some necessary 
information about theta-functions is given.

\section{The multi-component Pfaff-Toda and DKP hierarchies}

The multi-component Pfaff-Toda hierarchy was introduced in our recent
paper \cite{SZ25a}. (Its one-component
version was introduced and 
studied in detail by Takasaki in \cite{Takasaki09}.)
In the $N$-component Pfaff-Toda hierarchy, the independent variables are
the ``times'',
\beq\label{times2}
\begin{array}{l}
{\bf t}=\{{\bf t}_1, {\bf t}_2, \ldots , {\bf t}_N\}, \qquad
{\bf t}_{\alpha}=\{t_{\alpha , 1}, t_{\alpha , 2}, t_{\alpha , 3}, 
\ldots \, \},
\\ \\
\bar {\bf t}=\{\bar {\bf t}_1, \bar {\bf t}_2, \ldots , 
\bar {\bf t}_N\}, \qquad
\bar {\bf t}_{\alpha}=\{\bar t_{\alpha , 1}, \bar t_{\alpha , 2}, 
\bar t_{\alpha , 3}, \ldots \, \},
\end{array}
\qquad \alpha = 1, \ldots , N
\eeq
(in general complex numbers) and two finite sets of integer variables
$$
{\bf n}=\{n_1, \ldots , n_N\}, \quad
\bar {\bf n}=\{\bar n_1, \ldots , \bar n_N\}, \quad n_{\alpha}, \bar n_{\alpha}
\in \ZZ .
$$
(Hereafter, the bar under any variable does not mean
complex conjugation, so the variables with and without bar should be regarded
as independent ones.) 

However, it should be stressed that the sets of 
variables ${\bf n}$ and $\bar {\bf n}$
are not fully independent. Set
\beq\label{sums}
|{\bf n}|=\sum_{\gamma =1}^N n_{\gamma}, \qquad
|\bar {\bf n}|=\sum_{\gamma =1}^N \bar n_{\gamma},
\eeq
then $|{\bf n}|$ and $|\bar {\bf n}|$ are required to be of the same
parity (both odd or both even):
\beq\label{sums1}
|{\bf n}|-|\bar {\bf n}|\in 2\ZZ .
\eeq
We will refer to condition (\ref{sums1}) as the parity condition.

The universal
dependent variable is the tau-function
$\tau ({\bf n}, \bar {\bf n}, {\bf t},
\bar {\bf t})$, where it is assumed that the discrete variables are
subject to the parity condition. The tau-function satisfies an infinite
number of bilinear equations.
As we shall see, shifts of the discrete
variables in these equations are such that they respect the parity
condition. Therefore, there are two independent sectors: even (if
$|{\bf n}|$ and $|\bar {\bf n}|$ are both even) and odd (if
$|{\bf n}|$ and $|\bar {\bf n}|$ are both odd).
One can put $\tau ({\bf n}, \bar {\bf n}, {\bf t},
\bar {\bf t})=0$ if $|{\bf n}|$ and $|\bar {\bf n}|$ are of different
parities\footnote{This convention is also justified by the fermionic
representation of the hierarchy, see \cite{SZ25a}.}. 

\subsection{The integral bilinear functional equation}
\label{section:integral}

In order to write down the generating bilinear equation for the 
tau-function, we recall some standard notation.
Let ${\bf e}_{\gamma}$ be
the $N$-component vector whose $\gamma$th component is 1 and all others
are zero.
By
$[z]_{\gamma}$ we denote the set of times ${\bf t}$
such that ${\bf t}_{\mu}=\{0,0, \ldots \}$ if
$\mu \neq \gamma$, ${\bf t}_{\gamma}=
\{z, z^2/2, z^3/3, \ldots \}$ and 
\beq\label{f51a}
\epsilon_{\gamma}({\bf n})=(-1)^{n_{\gamma +1}+\ldots +n_N}
\eeq
is a sign factor. We will also use the notation
\beq\label{xi}
\xi ({\bf t}_{\gamma}, z)=\sum_{k\geq 1}t_{\gamma , k}z^k.
\eeq
In this notation, the equation for the tau-function
obtained in \cite{SZ25a} has the form
\beq\label{main}
\begin{array}{l}
\displaystyle{
\sum_{\gamma =1}^N\epsilon_{\gamma}({\bf n})\epsilon_{\gamma}({\bf n}')
\oint_{C_{\infty}}\frac{dz}{z^2} 
z^{n_{\gamma}-n_{\gamma}'}
e^{\xi ({\bf t}_{\gamma}-{\bf t}_{\gamma}', z)}}
\\ \\
\phantom{aaaaaaaaaaa}\displaystyle{\times \, 
\tau ({\bf n}\! -\! {\bf e}_{\gamma}, \bar {\bf n}, 
{\bf t}\! -\! [z^{-1}]_{\gamma},
\bar {\bf t})\tau ({\bf n}'\! +\! {\bf e}_{\gamma}, \bar {\bf n}', 
{\bf t}'\! +\! [z^{-1}]_{\gamma},
\bar {\bf t}')}
\\ \\
+\, 
\displaystyle{
\sum_{\gamma =1}^N \epsilon_{\gamma}({\bf n})\epsilon_{\gamma}({\bf n}')
\oint_{C_{\infty}}\frac{dz}{z^2} 
z^{n_{\gamma}'-n_{\gamma}}
e^{-\xi ({\bf t}_{\gamma}-{\bf t}_{\gamma}', z)}}
\\ \\
\phantom{aaaaaaaaaaa}\displaystyle{\times \, 
\tau ({\bf n}\! +\! {\bf e}_{\gamma}, \bar {\bf n}, 
{\bf t}\! +\! [z^{-1}]_{\gamma},
\bar {\bf t})\tau ({\bf n}'\! -\! {\bf e}_{\gamma}, \bar {\bf n}', 
{\bf t}'\! -\! [z^{-1}]_{\gamma},
\bar {\bf t}')}
\end{array}
\eeq
$$
\begin{array}{l}
=\, 
\displaystyle{
\sum_{\gamma =1}^N \epsilon_{\gamma}(\bar {\bf n})
\epsilon_{\gamma}(\bar {\bf n}')
\oint_{C_{\infty}}\frac{dz}{z^2} 
z^{\bar n_{\gamma}-\bar n_{\gamma}'}
e^{\xi (\bar {\bf t}_{\gamma}-\bar {\bf t}_{\gamma}', z)}}
\\ \\
\phantom{aaaaaaaaaaa}\displaystyle{\times \, 
\tau ({\bf n}, \bar {\bf n}\! -\! {\bf e}_{\gamma}, 
{\bf t},
\bar {\bf t}\! -\! [z^{-1}]_{\gamma})
\tau ({\bf n}', \bar {\bf n}'\! +\! {\bf e}_{\gamma}, 
{\bf t}',
\bar {\bf t}'\! +\! [z^{-1}]_{\gamma})}
\\ \\
+\, 
\displaystyle{
\sum_{\gamma =1}^N \epsilon_{\gamma}(\bar {\bf n})
\epsilon_{\gamma}(\bar {\bf n}')
\oint_{C_{\infty}}\frac{dz}{z^2} 
z^{\bar n_{\gamma}'-\bar n_{\gamma}}
e^{-\xi (\bar {\bf t}_{\gamma}-\bar {\bf t}_{\gamma}', z)}}
\\ \\
\phantom{aaaaaaaaaaa}\displaystyle{\times \, 
\tau ({\bf n}, \bar {\bf n}\! +\! {\bf e}_{\gamma}, 
{\bf t},
\bar {\bf t}\! +\! [z^{-1}]_{\gamma})
\tau ({\bf n}', \bar {\bf n}'\! -\! {\bf e}_{\gamma}, 
{\bf t}',
\bar {\bf t}'\! -\! [z^{-1}]_{\gamma})}.
\end{array}
$$
It is valid for all ${\bf t}, \bar {\bf t}$, ${\bf t}', \bar {\bf t}'$
and ${\bf n}, \bar {\bf n}$,  ${\bf n}', \bar {\bf n}'$
such that 
\beq\label{par1}
|{\bf n}|-|\bar {\bf n}|\in 2\ZZ +1, \qquad
|{\bf n}'|-|\bar {\bf n}'|\in 2\ZZ +1,
\eeq
otherwise the parity condition is not satisfied (the right-hand sides of
(\ref{sums1}) and (\ref{par1}) have different parities because the 
parity condition (\ref{sums1}) should be applied to 
$|{\bf n}\pm {\bf e}_{\gamma}|$ and $|\bar {\bf n}|$). 
So, in (\ref{main}) it is assumed that
the numbers $|{\bf n}|$ and $|\bar {\bf n}|$ as well as
$|{\bf n}'|$ and $|\bar {\bf n}'|$ have different 
parities.
The integration contour $C_{\infty}$ 
around $\infty$ is a big circle of radius
$R\to \infty$ such that all singularities 
coming from the powers of $z$ and the exponential functions
$e^{\xi ({\bf t}_{\gamma}-{\bf t}_{\gamma}', \, z)}$, 
$e^{\xi (\bar {\bf t}_{\gamma}-\bar {\bf t}_{\gamma}', \, z)}$
are outside it and all singularities 
coming from the $\tau$-factors are inside it (the 
$\tau$-factors as functions of $z$ are assumed to be regular in some
neighborhood of infinity.)
At $N=1$ equation (\ref{main}) coincides, after a linear change 
of the discrete variables) with
the equation for the tau-function of the one-component Pfaff-Toda
hierarchy obtained by Takasaki 
in \cite{Takasaki09}. 
Note that equation (\ref{main}) has the following obvious 
symmetries:
\beq\label{sym1}
({\bf n}, {\bf t}, \bar {\bf n}, \bar {\bf t}),
({\bf n}', {\bf t}', \bar {\bf n}', \bar {\bf t}')
\longleftrightarrow
({\bf n}', {\bf t}', \bar {\bf n}', \bar {\bf t}'),
({\bf n}, {\bf t}, \bar {\bf n}, \bar {\bf t}),
\eeq
and
\beq\label{sym2}
({\bf n}, {\bf t}, {\bf n}', {\bf t}'),
(\bar {\bf n},\bar {\bf t}, \bar {\bf n}', \bar {\bf t}')
\longleftrightarrow
(\bar {\bf n},\bar {\bf t}, \bar {\bf n}', \bar {\bf t}'),
({\bf n}, {\bf t}, {\bf n}', {\bf t}').
\eeq 

Note also that the transformation
\beq\label{trans}
\tau ({\bf n}, \bar {\bf n}, {\bf t}, \bar {\bf t})
\longrightarrow
e^{i\pi |{\bf n}||\bar {\bf n}| +L({\bf n}, \bar {\bf n}, 
{\bf t}, \bar {\bf t})}
\tau ({\bf n}, \bar {\bf n}, {\bf t}, \bar {\bf t}),
\eeq
where $L({\bf n}, \bar {\bf n}, {\bf t}, \bar {\bf t})$ is an arbitrary
linear form in the space spanned by the variables,
sends any solution to equation 
(\ref{par1}) to another solution. So, we can say that the tau-function
is defined by equation 
(\ref{par1}) up to a common multiplier as in (\ref{trans}). This remark
will be important below in section \ref{section:ellipticToda}.

\subsection{The $N$-component Pfaff-Toda as the $2N$-component DKP}
\label{section:TodaDKP}

For any $N\geq 1$, after setting 
$\bar {\bf n}'=\bar {\bf n}$, $\bar {\bf t}'=\bar {\bf t}$
in (\ref{main}) the bar-variables do not
participate in the equation entering as parameters. 
The right-hand side of (\ref{main}) vanishes identically
and the rest becomes the integral bilinear
equation for the tau-function of the $N$-component 
DKP hierarchy obtained in \cite{SZ24}. In this hierarchy,
the independent variables are ${\bf n}$ and ${\bf t}$, and
the tau-function will be denoted as
$\tau^{\rm DKP}({\bf n}, {\bf t})$.
For the case of $M$ components the equation reads:
\beq\label{main1}
\begin{array}{l}
\displaystyle{
\sum_{\gamma =1}^M\epsilon_{\gamma}({\bf n}\! -\! {\bf n}')
\oint_{C_{\infty}}\! \frac{dz}{z^2} 
z^{n_{\gamma}-n_{\gamma}'}
e^{\xi ({\bf t}_{\gamma}-{\bf t}_{\gamma}', z)}}
\\ \\
\phantom{aaaaaaaaaaaaa}
\displaystyle{
\times \tau^{\rm DKP} ({\bf n}\! -\! {\bf e}_{\gamma}, 
{\bf t}\! -\! [z^{-1}]_{\gamma})\tau^{\rm DKP} ({\bf n}'\! +\! {\bf e}_{\gamma}, 
{\bf t}'\! +\! [z^{-1}]_{\gamma})}
\\ \\
+\, 
\displaystyle{
\sum_{\gamma =1}^M \epsilon_{\gamma}({\bf n}\! -\! {\bf n}')
\oint_{C_{\infty}}\! \frac{dz}{z^2} 
z^{n_{\gamma}'-n_{\gamma}}
e^{-\xi ({\bf t}_{\gamma}-{\bf t}_{\gamma}', z)}}
\\ \\
\phantom{aaaaaaaaaaaaa}
\displaystyle{
\times \tau^{\rm DKP} ({\bf n}\! +\! {\bf e}_{\gamma}, 
{\bf t}\! +\! [z^{-1}]_{\gamma})\tau^{\rm DKP} ({\bf n}'\! -\! {\bf e}_{\gamma}, 
{\bf t}'\! -\! [z^{-1}]_{\gamma})}=0.
\end{array}
\eeq
It is valid for all ${\bf t}$, ${\bf t'}$ and ${\bf n}$, ${\bf n'}$ obeying
the parity condition which in this case means that $|{\bf n}|$ and
$|{\bf n'}|$ are both odd since $|{\bf n}|$ in 
$\tau^{\rm DKP}({\bf n}, {\bf t})$ 
should be even\footnote{In fact in \cite{SZ24} a little
bit less general version was dealt with: in that paper all components
of the vector ${\bf n}$ in $\tau^{\rm DKP}({\bf n}, {\bf t})$, 
not just their sum, were assumed to be even.}.
Note the obvious symmetry of this equation:
\beq\label{sym3}
({\bf n}, {\bf t}) \leftrightarrow ({\bf n'}, {\bf t'}).
\eeq

It is not difficult to see that the simplest solution to equation
(\ref{main1}) is
\beq\label{simplest}
\tau^{\rm DKP} ({\bf n}, {\bf t})=
\exp \left ( \frac{1}{2}\sum_{\gamma =1}^M \sum_{k\geq 1} kt_{\gamma , k}^2
\right ).
\eeq
Indeed,
plugging this into (\ref{main1}), we see, extracting 
a common multiplier, that the left-hand side
is proportional to
$$
\begin{array}{lll}
L&=&\displaystyle{\sum_{\gamma =1}^M\epsilon_{\gamma}({\bf n}-{\bf n'})
\oint_{C_{\infty}}\! \frac{dz}{z^2-1} \,
z^{n_{\gamma}-n_{\gamma}'}} e^{\xi ({\bf t}-{\bf t'}, z)-
\xi ({\bf t}-{\bf t'}, z^{-1})}
\\ && \\
&& \displaystyle{
+\, \sum_{\gamma =1}^M\epsilon_{\gamma}({\bf n}-{\bf n'})
\oint_{C_{\infty}}\! \frac{dz}{z^2-1} \, 
z^{n_{\gamma}'-n_{\gamma}}}e^{-\xi ({\bf t}-{\bf t'}, z)+
\xi ({\bf t}-{\bf t'}, z^{-1})}.
\end{array}
$$
Changing the integration variable $z\to z^{-1}$ in the second line,
we have:
$$
\begin{array}{lll}
L&=&\displaystyle{\sum_{\gamma =1}^M\epsilon_{\gamma}({\bf n})
\epsilon_{\gamma}({\bf n'})
\Bigl (\oint_{C_{\infty}}-\oint_{C_{0}}\Bigr ) \frac{dz}{z^2-1} \, 
z^{n_{\gamma}-n_{\gamma}'}} e^{\xi ({\bf t}-{\bf t'}, z)-
\xi ({\bf t}-{\bf t'}, z^{-1})},
\end{array}
$$
where $C_0$ is a small circle around $0$. Therefore, the integral is 
given by residues at the simple poles at the points $z=\pm 1$ lying inside
the annulus bordered by the circles $C_0$ and $C_{\infty}$:
$$
\begin{array}{lll}
L&=&\displaystyle{
\pi i\sum_{\gamma =1}^M\epsilon_{\gamma}({\bf n})
\epsilon_{\gamma}({\bf n'})\Bigl (1-(-1)^{n_{\gamma}-n'_{\gamma}}\Bigr )}
\\ && \\
&&=\, \displaystyle{\pi i\sum_{\gamma =1}^M \Bigl (\epsilon_{\gamma}({\bf n})
\epsilon_{\gamma}({\bf n'})-\epsilon_{\gamma -1}({\bf n})
\epsilon_{\gamma -1}({\bf n'})\Bigr )}
\\ && \\
&& =\, \pi i \, \Bigl (1-(-1)^{|{\bf n}|-|{\bf n'}|}\Bigr ) \, =\, 0
\end{array}
$$
because $|{\bf n}|$ and $|{\bf n'}|$ are both even numbers.

Our main goal in this section is to show that equations (\ref{main}) and
(\ref{main1}) are actually equivalent. More precisely, equation 
(\ref{main}) is equivalent to (\ref{main1}) at $M=2N$.
To see this, we re-denote the variables in (\ref{main1}) in 
the following special way. Let the index $\mu$ run from $1$ to $N$ and set
\beq\label{re}
n_{N+\mu}=\bar n_{\mu}, \qquad 
{\bf t}_{N+\mu}=\bar {\bf t}_{\mu}.
\eeq
Divide each sum over $\gamma$ in (\ref{main1}) in two: one from $1$ to $N$ and
the other from $N+1$ to $2N$. Then, after the obvious redefinition of the
tau-function, the first sum in (\ref{main1}) becomes
almost equal to the left-hand side of (\ref{main}) while the second one in
(\ref{main1}) is almost the right-hand side of (\ref{main}). ``Almost'' because
it still remains to identify the sign factors. Their comparison shows that
what comes from (\ref{main1}) as the left-hand side of (\ref{main}) contains
an extra sign factor $(-1)^{|\bar {\bf n}|-|\bar {\bf n}'|+1}$. As is easy
to see, it can be eliminated after multiplying the tau-function by 
the sign factor $(-1)^{\frac{1}{2}|\bar {\bf n}| (|\bar {\bf n}|-1)}$,
i.e., we identify
\beq\label{tautau}
\tau ({\bf n}, \bar {\bf n}, {\bf t}, \bar {\bf t})=
(-1)^{\frac{1}{2}|\bar {\bf n}| (|\bar {\bf n}|-1)}
\tau^{\rm DKP}(\tilde {\bf n}, \tilde {\bf t}),
\eeq
where the sets of the variables $\tilde {\bf n}, \tilde {\bf t}$ 
are $$\tilde {\bf n}=\{n_1, \ldots , n_N, \bar n_1, \ldots , \bar n_N\},
\quad
 \tilde {\bf t}=\{{\bf t}_1, \ldots , {\bf t}_N, 
 \bar {\bf t}_1, \ldots , \bar {\bf t}_N\}.$$
Note that the parity condition on the DKP side
is $|{\bf n}|+|\bar {\bf n}|\in 2\ZZ$,
while that on the Pfaff-Toda side is
$|{\bf n}|-|\bar {\bf n}|\in 2\ZZ$ which is the same.

Summarizing, we see that 
the relation between multi-component Pfaff-Toda and DKP
hierarchies is two-fold. On the one hand, the latter is a subhierarchy 
of the former (``a half'' of it). On the other hand, the $N$-component
Pfaff-Toda can be regarded, after simple re-definitions,
as the $2N$-component DKP. Below we will use this fact for analysis of the
dispersionless version of the Pfaff-Toda hierarchy.

\subsection{Bilinear equations of the Hirota-Miwa type}

Keeping in mind the result of section \ref{section:TodaDKP},
we hereafter deal with the DKP hierarchy for $M=2N$, i.e., with
equation (\ref{main1}), as it is much simpler than the original
version (\ref{main}) (two sums instead of four). At the end, one 
always can rewrite the results in the Toda variables using 
the identification (\ref{re}), (\ref{tautau}).
So, our independent variables are
\beq\label{times2a}
\begin{array}{l}
{\bf t}=\{{\bf t}_1, {\bf t}_2, \ldots , {\bf t}_M\}, \qquad
{\bf t}_{\alpha}=\{t_{\alpha , 1}, t_{\alpha , 2}, t_{\alpha , 3}, 
\ldots \, \},
\\ \\
{\bf n}=\{n_1, \ldots , n_M\}, \quad
\quad n_{\alpha} \in \ZZ , \quad \alpha = 1, \ldots , M,
\end{array}
\eeq
with the parity condition 
$$
|{\bf n}|\in 2\ZZ ,
$$
and the tau-function is $\tau ({\bf n}, {\bf t})$ (to simplify the notation,
we will write simply $\tau$, not $\tau^{\rm DKP}$, if this does not lead
to a misunderstanding). The general bilinear equation for it is (\ref{main1}).

To proceed, one should employ the Miwa change of
variables (first introduced by Miwa in \cite{Miwa82}).
As usual, passing to Miwa variables consists in choosing ${\bf t}-{\bf t}'$ 
in such a way that the integrals
could be evaluated by means of residue calculus. 
There are many different possibilities to do that. 

Fix some points $a_i, b_i \in \CC$ and put 
$\displaystyle{
{\bf t}-{\bf t}'=\sum_i [a_i^{-1}]_{\alpha_i}-
\sum_i [b_i^{-1}]_{\beta_i},}
$
assuming that the sums are finite. Here
$\alpha_i, \beta_i$ are arbitrary indices from the set
$\{1, \ldots , M\}$ (some of them may coincide). 
After this substitution, the integrals in (\ref{main1})
can be calculated by taking residues at poles at the points $a_i, b_i$
(the poles are simple if all of them are distinct). Besides, depending
on the choice of ${\bf n}-{\bf n}'$, there may be
a nonzero residue at infinity. Following the approach suggested in \cite{SZ25a},
we observe that for the particular
choice of ${\bf n}-{\bf n}'$ such that
each shift ${\bf t}_{\alpha}\to {\bf t}_{\alpha}\pm [a^{-1}]_{\alpha}$
is supplemented by the corresponding shift $n_{\alpha}\to n_{\alpha}
\pm 1$, the residue at infinity is always zero.

Keeping this in mind, we consider the following substitutions of
general form:
\beq\label{m1}
\begin{array}{l}
\displaystyle{
{\bf n}-{\bf n}' =-{\bf k}=\sum_{i=1}^{P^+}{\bf e}_{\alpha_i}-
\sum_{k=1}^{P^-}{\bf e}_{\beta_k},}
\\  \\
\displaystyle{
{\bf t}-{\bf t}' =-{\bf T}=\sum_{i=1}^{P^+}[a_i^{-1}]_{\alpha_i}-
\sum_{k=1}^{P^-}[b_k^{-1}]_{\beta_k}.}
\end{array}
\eeq
Here $P^+, P^-$ are non-negative integer numbers,
$\alpha_i, \beta_k$ are arbitrarily taken indices
from the set $\{1, \ldots , M\}$ (they may enter with multiplicities, i.e.,
the cases when $\alpha_i =\alpha_j$ or 
$\beta_i =\beta_j$ for $i\neq j$ are allowed), and
$a_i , b_k \in \CC$ are arbitrary 
parameters (the Miwa variables) 
belonging to a neighborhood of infinity (and, again, we allow the
cases when some of them coincide). 
If one or more points tend to infinity,
a nonzero residue at infinity arises. As is easy to see, the two procedures
(taking residue at some $a_i$ and tending $a_i$ to $\infty$,  
thus obtaining a non-zero residue there) commute.
Therefore, without any loss of generality, we can start from the case
when all complex numbers $a_i, b_k$ are finite and, if necessary, tend 
some of them to infinity afterwards.
It is important to note that the parity condition leads to the
following restriction on possible values of $P^{\pm}$:
\beq\label{m2}
P^+-P^-\in 2\ZZ .
\eeq
Following the terminology used in \cite{SZ25a},
we call the case when $a_i , b_k $  
and $\alpha_i, \beta_k$
are all distinct {\it non-degenerate}. In this case all poles are simple
and there are no contributions from infinity. 

The key formula necessary for reducing the integrals in 
(\ref{main1}) to finite sums of residues is simply 
\beq\label{m3}
z\, e^{\xi ([a^{-1}], z)}=z\, e^{-\log (1-z/a)}=\frac{z\, a}{a-z}=
\frac{1}{z^{-1}-a^{-1}}.
\eeq
For the substitution of the general form (\ref{m1}) 
the corresponding factor is
$$
z^{n_{\gamma}-n'_{\gamma}}\, e^{\xi ({\bf t}_{\gamma}-{\bf t}'_{\gamma}, z)}
=\prod_{i=1}^{P^+}(z^{-1}-a_i^{-1})^{-\delta_{\alpha_i \gamma}}
\prod_{k=1}^{P^-}(z^{-1}-b_k^{-1})^{\delta_{\beta_k \gamma}}.
$$
A similar factorization holds also for the $\epsilon$-factors:
\beq\label{m4}
\epsilon_{\gamma}({\bf n}- {\bf n}')=\prod_{i=1}^{P^+}
\epsilon_{\alpha_i \gamma}
\prod_{k=1}^{P^-}
\epsilon_{\beta_k \gamma},
\eeq
where the sign factor $\epsilon_{\alpha \beta}$ is defined as follows:
\beq\label{epsilon}
\epsilon_{\alpha \beta}=\left \{
\begin{array}{l}
\;\; 1 \quad \mbox{if $\alpha \leq \beta$},
\\ \\
-1 \quad \mbox{if $\alpha > \beta$}.
\end{array}
\right.
\eeq
It is convenient to introduce the function
\beq\label{m5}
E_{\alpha \beta}(a, b)=\epsilon_{\alpha \beta}
(a^{-1}-b^{-1})^{\delta_{\alpha \beta}}.
\eeq
Obviously, $E_{\beta \alpha}(b, a)=-E_{\alpha \beta}(a, b)$.

A direct calculation of residues in (\ref{main1}) after the substitution 
(\ref{m1}) leads to the following general Hirota-Miwa equation:
\beq\label{genHM}
\begin{array}{l}
\displaystyle{
\sum_{s=1}^{P^+}
\prod_{{\scriptsize \begin{array}{l}i=1\\ i\neq s \end{array}}}^{P^+}
E^{-1}_{\alpha_i \alpha_s}(a_s, a_i)
\prod_{k=1}^{P^-}
E_{\beta_k \alpha_s}(a_s, b_k)}
\\ \\
\phantom{aaaaaaa}\displaystyle{
\times \, 
\tau \Bigl ({\bf n}-{\bf e}_{\alpha_s},  {\bf t}-[a_s^{-1}]_{\alpha_s},\Bigr )
\tau \Bigl ({\bf n}+{\bf k}+{\bf e}_{\alpha_s}, 
{\bf t}+{\bf T}+[a_s^{-1}]_{\alpha_s},
\Bigr )}
\\ \\
\displaystyle{
+\sum_{s=1}^{P^-}
\prod_{{\scriptsize \begin{array}{l}k=1\\ k\neq s \end{array}}}^{P^-}
E^{-1}_{\beta_k \beta_s}(b_s, b_k)
\prod_{i=1}^{P^+}
E_{\alpha_i \beta_s}(b_s, a_i)}
\\ \\
\phantom{aaaaaaa}\displaystyle{
\times \, 
\tau \Bigl ({\bf n}+{\bf e}_{\beta_s}, 
{\bf t}+[b_s^{-1}]_{\beta_s}, \Bigr )
\tau \Bigl ({\bf n}+{\bf k}-{\bf e}_{\beta_s},  
{\bf t}+{\bf T}-[b_s^{-1}]_{\beta_s},
\Bigr )=0,}
\end{array}
\eeq
where ${\bf k}$ and ${\bf T}$ are given in (\ref{m1}).

In order to represent the equation in a more compact form,
one can use the short-hand notation for the independent
discrete variables introduced in \cite{SZ25a}:
\beq\label{nd6}
{\bf n}^{\alpha}={\bf n} +{\bf e}_{\alpha}, 
\quad
{\bf n}^{\alpha \beta}={\bf n} +{\bf e}_{\alpha}+{\bf e}_{\beta}
\eeq
and so on. Similar notations will be used for shifts of continuous
times:
\beq\label{nd6a}
{\bf t}^{[a_{\alpha}]}={\bf t} + [a^{-1}]_{\alpha}, 
\quad
{\bf t}^{[a_{\alpha}b_{\beta}]}={\bf t} + [a^{-1}]_{\alpha}+ [b^{-1}]_{\beta},
\eeq
and so on. 
For example,
$$
\tau ({\bf n}+{\bf e}_{\alpha}, {\bf t}+[a^{-1}]_{\alpha}+ [b^{-1}]_{\beta})=
\tau ({\bf n}^{\alpha},  {\bf t}^{[a_{\alpha}b_{\beta}]}).
$$

In this notation (and after some sifts of the variables), equation
(\ref{genHM}) acquires the form
\beq\label{genHM1}
\begin{array}{l}
\displaystyle{
\sum_{s=1}^{P^+}
\prod_{{\scriptsize \begin{array}{l}i=1\\ i\neq s \end{array}}}^{P^+}
E^{-1}_{\alpha_i \alpha_s}(a_s, a_i)
\prod_{k=1}^{P^-}
E_{\beta_k \alpha_s}(a_s, b_k)}
\\ \\
\phantom{aaaaaaa}\displaystyle{
\times \, 
\tau \Bigl ({\bf n}^{\alpha_1 \ldots \hat \alpha_s \ldots \alpha_{P^+}},  
{\bf t}^{[a_1\ldots \hat a_s \ldots a_{P^+}]}\Bigr )
\tau \Bigl ({\bf n}^{\alpha_s \beta_1 \ldots \beta_{P^-}}, 
{\bf t}^{[a_s b_1 \ldots b_{P^-}]} \Bigr )}
\\ \\
\displaystyle{
+\sum_{s=1}^{P^-}
\prod_{{\scriptsize \begin{array}{l}k=1\\ k\neq s \end{array}}}^{P^-}
E^{-1}_{\beta_k \beta_s}(b_s, b_k)
\prod_{i=1}^{P^+}
E_{\alpha_i \beta_s}(b_s, a_i)}
\\ \\
\phantom{aaaaaaa}\displaystyle{
\times \, 
\tau \Bigl ({\bf n}^{\beta_s \alpha_1 \ldots \alpha_{P^+}},  
{\bf t}^{[b_s a_1 \ldots a_{P^+}]}\Bigr )
\tau \Bigl ({\bf n}^{\beta_1 \ldots \hat \beta_s \ldots \beta_{P^-}}, 
{\bf t}^{[b_1 \ldots \hat b_s \ldots b_{P^-}]} \Bigr )}
=0,
\end{array}
\eeq
where hat above any letter means that this symbol should be omitted.
Equation (\ref{genHM1}) 
is the most general non-degenerate Hirota-Miwa equation
(functional relation) for the tau-function. 
It contains 
$$
P =P^+ +P^- 
$$
bilinear terms, each of which is product of two tau-functions with
various shifts of the arguments. The coefficients are rational functions
of $a_i , b_k$. We will call it
the (non-degenerate) $P$-point relation, according to the total
number of points. 
So, the number of terms in non-degenerate
relations always coincides with the number of points. 
Besides, in the non-degenerate 
case, when all the points are distinct, the relation does not contain any
derivatives of the tau-function with respect to the continuous times.
Note, however, that this general $P$-point relation still holds in 
degenerate cases, when some of the points merge or tend to infinity.
In such cases, some terms of the general equation (\ref{genHM}) 
become singular, if one considers them separately. However, in the full
expression, taking into account all the terms, the singularities 
can be resolved, and, as a result, 
derivatives of the tau-functions with respect to 
continuous times may arise. 

Note also that if some points in the non-degenerate $P$-point relation 
tend to infinity, the number
of points (i.e., the number of the remaining free parameters), $p$, 
in such degenerate relations becomes 
strictly less than $P$. In this case 
we will call them, following \cite{SZ25a}, $p$-point relations (not
necessarily non-degenerate), and consider them as reductions of some
non-degenerate relations for some $P> p$. (The number of
different bilinear terms in them is still $P$.) 
In the next subsection we present some details
of the simplest and most important particular case $P=4$.

\subsection{Non-degenerate 4-point relations}
\label{subsection:4-point}
 
In this section we take
a closer look at non-degenerate 4-point relations. As we shall see,
their classification is very simple:
there are only two essentially different types 
of them\footnote{This should be compared with a whole zoo of 2-point
(degenerate) relations obtained in \cite{SZ24}. In fact
all of them can be obtained from just two non-degenerate 4-point 
relations as a result of various degeneration procedures.}.
Here we do not consider 4-point relations that can be
obtained as degenerations of $P$-point ones for $P>4$ because they 
contain more than four terms.

First of all, consider the simplest case $P=2$.
It turns out to be
trivial: it is easy to check that each non-degenerate $2$-point relation 
contains just 2 terms and converts into identity. Since $P$ must be even,
non-degenerate 3-point 
relations in this hierarchy do not exist, and 
any 2- or 3-point Hirota-Miwa relation containing not more than four
terms is a reduction of some non-degenerate 4-point relation.

The case of our main interest is $P=4$: 
\beq\label{nd1}
P^+ + P^- =4.
\eeq
There are five possibilities but taking into account the symmetry 
(\ref{sym3}) we can consider only the following three:
\beq\label{nd1a}
(P^+, P^-)=(4,0), \; (3,1), \; (2,2).
\eeq
Let us write down the corresponding Hirota-Miwa 
equations explicitly. For the notational simplicity 
we use the notation $a,b,c,d\in \CC$ for the four points and 
$\alpha , \beta , \nu , \mu \in \{1, \ldots , M\}$ for the four 
corresponding indices\footnote{Each index is ``linked'' to the corresponding
point: $\alpha$ to $a$, $\beta$ to $b$, $\nu$ to $c$, $\mu$ to $d$.}.
Actually, some of them may coincide but we are interested in the general 
non-degenerate case
assuming that all of them are distinct.

For the first possibility, 
$(P^ +, P^-)=(4,0)$, we have:
\beq\label{40}
\begin{array}{l}
{\bf n}-{\bf n}'={\bf e}_{\alpha}+{\bf e}_{\beta}+{\bf e}_{\nu}+{\bf e}_{\mu}, 
\\  \\
{\bf t}-{\bf t}'=[a^{-1}]_{\alpha} +[b^{-1}]_{\beta}+ [c^{-1}]_{\nu}
+ [d^{-1}]_{\mu}.
\end{array}
\eeq
Calculating the residues, or specializing the general 
equation (\ref{genHM}) to this particular case,
we can represent, after some simple transformations (like
shifts of the variables), the corresponding Hirota-Miwa equation in the form
\beq\label{40a}
\begin{array}{l}
\phantom{-}E_{\nu \beta}(b,c)E_{\mu \beta}(b,d)E_{\nu \mu}(d,c)
\tau \Bigl ({\bf n}^{\beta \nu \mu}, {\bf t}^{[b_{\beta}c_{\nu}d_{\mu}]}\Bigr )
\tau \Bigl ({\bf n}^{\alpha}, {\bf t}^{[a_{\alpha}]}\Bigr )
\\ \\
-\, E_{\nu \alpha}(a,c)E_{\mu \alpha}(a,d)E_{\nu \mu}(d,c)
\tau \Bigl ({\bf n}^{\alpha \nu \mu}, {\bf t}^{[a_{\alpha}c_{\nu}d_{\mu}]}\Bigr )
\tau \Bigl ({\bf n}^{\beta}, {\bf t}^{[b_{\beta}]}\Bigr )
\\ \\
-\, E_{\beta \alpha}(a,b)E_{\mu \alpha}(a,d)E_{\mu \beta}(b,d)
\tau \Bigl ({\bf n}^{\alpha \beta \mu}, {\bf t}^{[a_{\alpha}b_{\beta}
d_{\mu}]}\Bigr )
\tau \Bigl ({\bf n}^{\nu}, {\bf t}^{[c_{\nu}]}\Bigr )
\\ \\
+\, E_{\beta \alpha}(a,b)E_{\nu \alpha}(a,c)E_{\nu \beta}(b,c)
\tau \Bigl ({\bf n}^{\alpha \beta \nu}, {\bf t}^{[a_{\alpha}b_{\beta}
c_{\nu}]}\Bigr )
\tau \Bigl ({\bf n}^{\mu}, {\bf t}^{[d_{\mu}]}\Bigr )\, =0.
\end{array}
\eeq
For the second possibility, 
$(P^ +, P^-)=(3,1)$, we have:
\beq\label{31}
\begin{array}{l}
{\bf n}-{\bf n}'={\bf e}_{\alpha}+{\bf e}_{\beta}+{\bf e}_{\nu}-{\bf e}_{\mu}, 
\\  \\
{\bf t}-{\bf t}'=[a^{-1}]_{\alpha} +[b^{-1}]_{\beta}+ [c^{-1}]_{\nu}-
[d^{-1}]_{\mu}.
\end{array}
\eeq
The corresponding Hirota-Miwa equation is
\beq\label{31a}
\begin{array}{l}
\phantom{-}E_{\nu \beta}(b,c)E_{\mu \alpha}(a,d)
\tau \Bigl ({\bf n}^{\beta \nu}, {\bf t}^{[b_{\beta}c_{\nu}]}\Bigr )
\tau \Bigl ({\bf n}^{\alpha \mu}, {\bf t}^{[a_{\alpha}d_{\mu}]}\Bigr )
\\ \\
-\, E_{\mu \beta}(b,d)E_{\nu \alpha}(a,c)
\tau \Bigl ({\bf n}^{\alpha \nu}, {\bf t}^{[a_{\alpha}c_{\nu}]}\Bigr )
\tau \Bigl ({\bf n}^{\beta \mu}, {\bf t}^{[b_{\beta}d_{\mu}]}\Bigr )
\\ \\
+\, E_{\beta \alpha}(a,b)E_{\mu \nu}(c,d)
\tau \Bigl ({\bf n}^{\alpha \beta}, {\bf t}^{[a_{\alpha}b_{\beta}]}\Bigr )
\tau \Bigl ({\bf n}^{\nu \mu}, {\bf t}^{[c_{\nu}d_{\mu}]}\Bigr )
\\ \\
+\, E_{\beta \alpha}(a,b)E_{\nu \alpha}(a,c)E_{\nu \beta}(b,c)
E_{\alpha \mu}(d,a)E_{\beta \mu}(d,b)E_{\nu\mu}(d,c)
\\ \\
\phantom{aaaaaaaaaaaaaaaaaaaaaa}
\times \tau \Bigl ({\bf n}^{\alpha \beta \nu \mu}, 
{\bf t}^{[a_{\alpha}b_{\beta}c_{\nu}d_{\mu}]}\Bigr )
\tau \Bigl ({\bf n}, {\bf t}\Bigr )\, =0.
\end{array}
\eeq
For the last possibility, 
$(P^ +, P^-)=(2,2)$, we have:
\beq\label{22}
\begin{array}{l}
{\bf n}-{\bf n}'={\bf e}_{\alpha}+{\bf e}_{\beta}-{\bf e}_{\nu}-{\bf e}_{\mu}, 
\\  \\
{\bf t}-{\bf t}'=[a^{-1}]_{\alpha} +[b^{-1}]_{\beta}- [c^{-1}]_{\nu}-
[d^{-1}]_{\mu}.
\end{array}
\eeq
The Hirota-Miwa equation obtained in this case turns out to be
equivalent to (\ref{40a}). Therefore, there are only two essentially
different non-degenerate 4-point relations: (\ref{40a}) and (\ref{31a}).

Recall that in our recent paper \cite{SZ25a} four (rather than two) 
different non-degenerate
4-point relations for the multi-component Pfaff-Toda have been found.
In the notation adopted in that paper they correspond to the schemes
$$
(L^+, L^- | R^+, R^-)= (3,0|1,0), \, (3,0|0,1), \, (2,0|2,0), \,
(2,0 |1,1)
$$
(equations (4.15), (4.17), (4.21) and (4.23) in \cite{SZ25a}, respectively).
It is easy to see that all of them are actually contained in (\ref{40a}),
(\ref{31a}) and can be obtained from them using different possibilities 
to divide the four Miwa variables in DKP in two groups of 
variables with and without bar. More precisely, the interrelations between
the equations is as follows:
$$
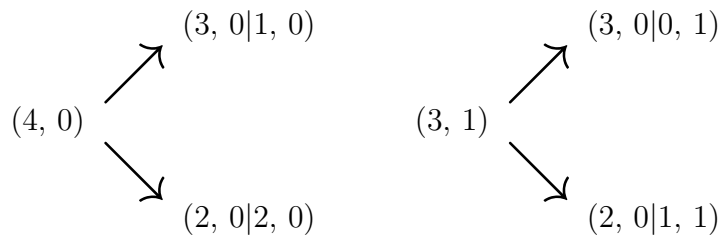
\begin{array}{lll}
& & \!\! (3, \, 0 | 1, \, 0)\\
&  \mbox{{\huge $\nearrow$}} & \\
\,\, (4, \, 0)\!\! &&\\
& \mbox{{\huge $\searrow$}} &\\
&& \!\! (2, \, 0 | 2, \, 0)
\end{array}
\qquad
\begin{array}{lll}
& & \!\! (3, \, 0 | 0, \, 1)\\
&  \mbox{{\huge $\nearrow$}} & \\
\,\, (3, \, 1)\!\! &&\\
& \mbox{{\huge $\searrow$}} &\\
&& \!\! (2, \, 0 | 1, \, 1)
\end{array}
$$
By performing different possible degenerations of the two equations (\ref{40a}),
(\ref{31a}) (such as tending two points to $\infty$ or merging two
or more points), one can reproduce all the 2-point equations listed 
in \cite{SZ24}. We will not discuss the details here.

\section{The dispersionless limit}

The standard reference to the dispersionless limits of integrable
hierarchies is \cite{TT95}.
In order to perform this limit for the 
multi-component Pfaff-Toda hierarchy, one should introduce a small
parameter $\hbar$ (which eventually is put equal to 0)
and re-scale the times ${\bf t}$, $\bar {\bf t}$
and the discrete variables ${\bf n}$, $\bar {\bf n}$ as
$t_{\alpha , k}\to t_{\alpha , k}/\hbar$, 
$n_{\alpha}\to t_{\alpha , 0}/\hbar$ and similarly 
for their bar-counterparts. Introduce the function 
$F({\bf t}_0, \bar {\bf t}_0,  {\bf t}, \bar {\bf t}; 
\hbar )$ related to the tau-function of the multi-component
Pfaff-Toda hierarchy by the
formula
\beq\label{d1}
\tau \Bigl (\hbar^{-1}{\bf t}_0 , \hbar^{-1}\bar {\bf t}_0 , 
\hbar^{-1}{\bf t} , \hbar^{-1}\bar {\bf t} \Bigr )=
\exp \left ( \frac{1}{\hbar^2}\,
F({\bf t}_0, \bar {\bf t}_0, {\bf t},\bar {\bf t}; \hbar )\right )
\eeq
and consider the limit  
$$
F=F({\bf t}_0, \bar {\bf t}_0, {\bf t},\bar {\bf t})=
\lim\limits_{\hbar \to 0}
F({\bf t}_0, \bar {\bf t}_0, {\bf t},\bar {\bf t}; \hbar ),$$ 
if it exists\footnote{This limit 
does exist, for example, for tau-functions that appear as
partition functions of random matrix models. In this context
the $F$-function is the free energy. However, for some classes
of tau-functions (for example, for soliton or algebro-geometric
solutions) this limit does not exist.}. 
The function $F$
plays the role of the 
tau-function in the dispersionless limit $\hbar \to 0$, meaning that
it serves as a universal dependent variable. (The tau-function itself
does not exist at $\hbar =0$.)
The $F$-function satisfies 
an infinite number of highly nonlinear differential equations which are
limiting cases of
the bilinear equations for the tau-function rewritten in terms of the
$F$-function. 
Note that the former
discrete variables ${\bf n}, \bar {\bf n}$ become continuous, and the
former difference equations containing these variables 
become differential.

Our main goal in this paper is to obtain the dispersionless version
of the multi-component Pfaff-Toda hierarchy and represent it in the
elliptic form. To this end, we will employ 
the Pfaff-Toda-DKP equivalence 
established in section
\ref{section:TodaDKP}. Namely, we are going to revisit the dispersionless
version of the multi-component DKP (earlier obtained in our paper \cite{SZ24}),
starting from the non-degenerate 4-point relations. This approach turns
out to be simultaneously more rigorous and much easier than that of 
\cite{SZ24} based on 2-point relations. At the final stage, when the 
results, including the elliptic uniformization are obtained, one should 
just rewrite it in terms of the Toda variables (with and without bar).

\subsection{Dispersionless version of the Hirota-Miwa equations}

To find the dispersionless limit of the Hirota-Miwa equations,
the first step is rewriting them
in terms of the $F$-function for any finite $\hbar$ in the form which
is most convenient for taking the limit $\hbar \to 0$. 
To this end,
we introduce the differential operator (or rather vector field in the
space of times)
\beq\label{n1}
\nabla_{\alpha}(z)= \p_{t_{\alpha , 0}} +\sum_{k\geq 1}\frac{z^{-k}}{k}\,
\p_{t_{\alpha , k}}.
\eeq
Then we can write:
$$
\tau \Bigl ({\bf t}_0/\hbar \pm {\bf e}_{\alpha} , 
{\bf t}/\hbar \pm [z^{-1}]_{\alpha}\Bigr )
=\exp \Bigl ( \frac{1}{\hbar^2}
e^{\pm \hbar \nabla_{\alpha}(z)}
F({\bf t}_0, {\bf t}; \hbar )\Bigr ).
$$

We begin with
the simplest nontrivial case, the 4-point relations
(\ref{40a}), (\ref{31a}). 
Equation (\ref{40a}) can be rewritten in the form
\beq\label{40b}
\begin{array}{l}
\phantom{-}E_{\nu \beta}(b,c)E_{\mu \beta}(b,d)E_{\nu \mu}(d,c)
\exp \left [\frac{1}{\hbar^2}\Bigl (
e^{\hbar (\nabla_{\beta}(b)+\nabla_{\nu}(c)+
\nabla_{\mu}(d)} +e^{\hbar (\nabla_{\alpha}(a)} \Bigr )F \right ]
\\ \\
-\, E_{\nu \alpha}(a,c)E_{\mu \alpha}(a,d)E_{\nu \mu}(d,c)
\exp \left [\frac{1}{\hbar^2}\Bigl (
e^{\hbar (\nabla_{\alpha}(a)+\nabla_{\nu}(c)+
\nabla_{\mu}(d)} +e^{\hbar (\nabla_{\beta}(b)} \Bigr )F \right ]
\\ \\
-\, E_{\beta \alpha}(a,b)E_{\mu \alpha}(a,d)E_{\mu \beta}(b,d)
\exp \left [\frac{1}{\hbar^2}\Bigl (
e^{\hbar (\nabla_{\alpha}(a)+\nabla_{\beta}(b)+
\nabla_{\mu}(d)} +e^{\hbar (\nabla_{\nu}(c)} \Bigr )F \right ]
\\ \\
+\, E_{\beta \alpha}(a,b)E_{\nu \alpha}(a,c)E_{\nu \beta}(b,c)
\exp \left [\frac{1}{\hbar^2}\Bigl (
e^{\hbar (\nabla_{\alpha}(a)+\nabla_{\beta}(b)+
\nabla_{\nu}(c)} +e^{\hbar (\nabla_{\mu}(d)} \Bigr )F \right ]
\, =0,
\end{array}
\eeq
from which the $\hbar \to 0$ limit is straightforward. The result is:
\beq\label{40c}
\begin{array}{l}
\phantom{-}E_{\nu \beta}(b,c)E_{\mu \beta}(b,d)E_{\nu \mu}(d,c)
e^{\bigl ( \nabla_{\beta}(b) \nabla_{\nu}(c)+ \nabla_{\beta}(b)
\nabla_{\mu}(d)+ \nabla_{\nu}(c)\nabla_{\mu}(d)\bigr )F}
\\ \\
-\, E_{\nu \alpha}(a,c)E_{\mu \alpha}(a,d)E_{\nu \mu}(d,c)
e^{\bigl ( \nabla_{\alpha}(a) \nabla_{\nu}(c)+ \nabla_{\alpha}(a)
\nabla_{\mu}(d)+ \nabla_{\nu}(c)\nabla_{\mu}(d)\bigr )F}
\\ \\
-\, E_{\beta \alpha}(a,b)E_{\mu \alpha}(a,d)E_{\mu \beta}(b,d)
e^{\bigl ( \nabla_{\alpha}(a) \nabla_{\beta}(b)+ \nabla_{\alpha}(a)
\nabla_{\mu}(d)+ \nabla_{\beta}(b)\nabla_{\mu}(d)\bigr )F}
\\ \\
+\, E_{\beta \alpha}(a,b)E_{\nu \alpha}(a,c)E_{\nu \beta}(b,c)
e^{\bigl ( \nabla_{\alpha}(a) \nabla_{\beta}(b )+ \nabla_{\alpha}(a)
\nabla_{\nu}(c)+ \nabla_{\beta}(b)\nabla_{\nu}(c)\bigr )F}
\, =0.
\end{array}
\eeq
In a similar way, one can take the limit of equation (\ref{31a}); the result is:
\beq\label{31c}
\begin{array}{l}
\phantom{-}E_{\nu \beta}(b,c)E_{\mu \alpha}(a,d)
e^{\bigl ( \nabla_{\beta}(b) \nabla_{\nu}(c)+ \nabla_{\alpha}(a)
\nabla_{\mu}(d)\bigr )F}
\\ \\
-\, E_{\mu \beta}(b,d)E_{\nu \alpha}(a,c)
e^{\bigl ( \nabla_{\alpha}(a) \nabla_{\nu}(c)+ \nabla_{\beta}(b)
\nabla_{\mu}(d\bigr )F}
\\ \\
+\, E_{\beta \alpha}(a,b)E_{\mu \nu}(c,d)
e^{\bigl ( \nabla_{\alpha}(a) \nabla_{\beta}(b)+ \nabla_{\nu}(c )
\nabla_{\mu}(d)\bigr )F}
\\ \\
+\, E_{\beta \alpha}(a,b)E_{\nu \alpha}(a,c)E_{\nu \beta}(b,c)
E_{\alpha \mu}(d,a)E_{\beta \mu}(d,b)E_{\nu\mu}(d,c)
\\ \\
\times 
e^{\bigl ( \nabla_{\alpha}(a) \nabla_{\beta}(b)+ \nabla_{\alpha}(a )
\nabla_{\nu}(c)+ \nabla_{\alpha}(a) \nabla_{\mu}(d)+
\nabla_{\beta}(b) \nabla_{\nu}(c)+ \nabla_{\beta}(b) \nabla_{\mu}(d)+
\nabla_{\nu}(c) \nabla_{\mu}(d)\bigr )F}
\, =0.
\end{array}
\eeq
The $E$-function is given by (\ref{m5}).

In the general case of arbitrary $P\geq 4$, 
the procedure is basically the same. After 
extracting a common factor, one obtains from (\ref{genHM1}):
\beq\label{genHM2}
\begin{array}{l}
\displaystyle{
\sum_{s=1}^{P^+}
\prod_{{\scriptsize \begin{array}{l}i=1\\ i\neq s \end{array}}}^{P^+}
E^{-1}_{\alpha_i \alpha_s}(a_s, a_i)
\prod_{k=1}^{P^-}
E_{\beta_k \alpha_s}(a_s, b_k)
\exp \!
\left (\nabla_{\alpha_s}(a_s) \Bigl (\sum_{k=1}^{P^-}\nabla_{\beta_k}(b_k)
-  \sum_{i\neq s}^{P^+}\nabla_{\alpha_i}(a_i)\Bigr )\right )
}
\\ \\
\displaystyle{
+\sum_{s=1}^{P^-}
\prod_{{\scriptsize \begin{array}{l}k=1\\ k\neq s \end{array}}}^{P^-}
E^{-1}_{\beta_k \beta_s}(b_s, b_k)
\prod_{i=1}^{P^+}
E_{\alpha_i \beta_s}(b_s, a_i)
\exp \! \left (\nabla_{\beta_s}(b_s) \Bigl 
(\sum_{i=1}^{P^+}\nabla_{\alpha_i}(a_i)
-  \sum_{k\neq s}^{P^-}\nabla_{\beta_k}(b_k)\Bigr )\right )
}
=0.
\end{array}
\eeq
Recall that $P=P^++P^-\geq 2$ here should be even.

\subsection{Elliptic uniformization of the multi-component DKP hierarchy
revisited}
\label{section:elliptic}

In \cite{SZ24} we have shown that all equations of the 
multi-component DKP hierarchy in the zero dispersion limit, in the elliptic
parametrization, 
are encoded in the single equation
\beq\label{E1}
E_{\beta \alpha}(a,b) e^{\nabla_{\alpha}(a)\nabla_{\beta}(b)F}
= \frac{\theta_1(u_{\alpha}(a)- u_{\beta}(b))}{\theta_4(u_{\alpha}(a)-
u_{\beta}(b))},
\eeq
where $\theta_1 (u), \, \theta_4(u)$ are Jacobi's theta-functions 
with an elliptic modular parameter $\tau$ not shown explicitly 
in (\ref{E1}) (see the Appendix).
In the right-hand side, $u_{\alpha}(z)=u_{\alpha}(z; {\bf t})$
is understood as a generating function of the form
\beq\label{u}
u_{\alpha}(z)=\eta_{\alpha}({\bf t}) + 
\sum_{k\geq 1}c^{(\alpha )}_k({\bf t}) z^{-k},
\qquad z\to \infty ,
\eeq
for  $\eta_{\alpha}({\bf t})$ and $c^{(\alpha )}_k({\bf t})$, 
which play the role of dynamical
variables. The meaning of equation (\ref{E1}) 
is that general second order derivatives
of the function $F$ with respect to the independent variables 
are expressed through
some special second order derivatives, as is usual in the 
theory of dispersionless equations. 

Let us explain this
in more details. For brevity, denote the function 
$\theta_1(u)/\theta_4(u)$ by $\mbox{sn}(u)$ and the inverse function
by $\mbox{arcsn}(u)$\footnote{Strictly speaking, 
the function $\mbox{sn}$ defined in this way is not quite what is called
the elliptic sinus function but is very similar to it: it differs 
from it by
a common $u$-independent factor and a re-scaling of the variable.}. 
Putting $\beta =\alpha$ and $b=\infty$ in (\ref{main}), we 
conclude that
$$
u_{\alpha}(a)-\eta_{\alpha}=
\mbox{arcsn} \Bigl (a^{-1}e^{\nabla_{\alpha}(a)
\p_{\alpha}F}\Bigr ),
$$
where $\p_{\alpha}\equiv \p_{t_{\alpha , 0}}$.
Thus, equation (\ref{main}) for $\alpha
\neq \beta$ can be written as
\beq\label{t16}
\epsilon_{\beta \alpha}e^{\nabla_{\alpha}(a)\nabla_{\beta}(b)F}
=\mbox{sn} \Bigl ( \eta_{\alpha}-\eta_{\beta}+\mbox{arcsn}\, (a^{-1}
e^{\nabla_{\alpha}(a)\p_{\alpha}F})-
\mbox{arcsn}\, (b^{-1}
e^{\nabla_{\beta}(b)\p_{\beta}F})\Bigr ).
\eeq
At the same time, since
$$
\eta_{\alpha}-\eta_{\beta} = \left \{
\begin{array}{l}
\displaystyle{\phantom{-}\sum_{\gamma =\alpha}^{\beta -1} 
(\eta_{\gamma} -\eta_{\gamma +1}),} \quad \alpha <\beta ,
\\  \\
\displaystyle{
-\sum_{\gamma =\beta}^{\alpha -1} (\eta_{\gamma}-\eta_{\gamma +1}),}  
\quad \alpha >\beta , 
\end{array} \right.
$$
we can write:
$$
\eta_{\alpha}-\eta_{\beta}= \left \{
\begin{array}{l}
\displaystyle{
-\sum_{\gamma =\alpha}^{\beta -1} 
\mbox{arcsn} (e^{\p_{\gamma}\p_{\gamma +1}F}),}
\quad \alpha <\beta ,
\\  \\
\displaystyle{ \phantom{-}
\sum_{\gamma =\beta}^{\alpha -1} 
\mbox{arcsn} (e^{\p_{\gamma}\p_{\gamma +1}F}),}
\quad \alpha >\beta .
\end{array} \right.
$$
Therefore, as equation (\ref{t16}) shows, the general 
second order derivatives of the function $F$ are expressed through the
particular derivatives $\p_{\alpha}^2 F$, $\p_{\alpha}\p_{\alpha +1}F$,
$\p_{t_{\alpha ,k}}\p_{\alpha}F$ for $\alpha =1, \ldots , M$. 

Putting $b=\infty$ in (\ref{E1}) and denoting $a=z$, we get
\beq\label{E1a}
\epsilon_{\beta \alpha}z^{-\delta_{\alpha \beta} }
e^{\nabla_{\alpha}(z)\p_{\beta}F}=
\frac{\theta_1(u_{\alpha}(z)-\eta_{\beta})}{\theta_4(u_{\alpha}(z)-\eta_{\beta})}.
\eeq
It is convenient to denote
\beq\label{E5}
R_{\alpha}=e^{\p_{\alpha}^2F}, \qquad
R_{\alpha \beta}=e^{\p_{\alpha} \p_{\beta}F}  =R_{\beta \alpha}\;\; (\mbox{for}\; 
\alpha \neq \beta ).
\eeq
The further limit $z\to \infty$ in (\ref{E1a}) yields, for $\beta \neq \alpha$:
\beq\label{E1b}
R_{\alpha \beta}=\epsilon_{\beta \alpha}
\frac{\theta_1(\eta_{\alpha \beta})}{\theta_4(\eta_{\alpha \beta})},
\qquad \eta_{\alpha \beta}\equiv \eta_{\alpha}-\eta_{\beta},
\eeq
which is explicitly symmetric in $\alpha$ and $\beta$, as it should be
according to the definition (\ref{E5}).
In the limit $z\to \infty$ in (\ref{E1a}) for $\beta =\alpha$ both sides
tend to zero. Comparison of the coefficients in front of $z^{-1}$ in the 
expansion around $\infty$ gives:
\beq\label{E1c}
R_{\alpha}=\pi c_1^{(\alpha )}\theta_2(0) \theta_3(0),
\eeq
where $c_1^{(\alpha )}$ is the coefficient at $z^{-1}$ in the expansion
(\ref{u}) (to obtain the right-hand side in this form, one should
use identity (\ref{theta1prime}) from the appendix).

It is convenient to introduce the function
\beq\label{S}
S(u)=\log \frac{\theta_1(u)}{\theta_4(u)}.
\eeq
It has the (quasi)periodicity properties $S(u+1)=S(u)+i\pi$,
$S(u+\tau )=S(u)$. Its $u$-derivative is already a double-periodic
function. It is given by\footnote{This formula can be proved comparing
the analytic properties of the two sides and using (\ref{theta1prime}).}
\beq\label{S'}
S'(u)=\pi \theta_4^2(0)\frac{\theta_2(u)\, \theta_3 (u)}{\theta_1(u)\, 
\theta_4(u)}.
\eeq
Note that identity (\ref{id}) from the appendix can be viewed as a nonlinear
differential equation for the function $S(u)$:
\beq\label{S1}
\left (\frac{S'(u)}{\pi \theta_2(0)\, \theta_3 (0)}\right )^2
=2\cosh (2S(u)) -\frac{\theta_2^2(0)}{\theta_3^2(0)}-
\frac{\theta_3^2(0)}{\theta_2^2(0)}.
\eeq

Now let us compare the next-to-leading terms, as $z\to \infty$, 
in the both
sides (\ref{E1a}) for $\alpha \neq \beta$. This gives:
$$
R_{\alpha \beta}\Bigl (1+\frac{c_1^{(\alpha )}}{z} S'(\eta_{\alpha \beta})
\Bigr ) =R_{\alpha \beta} \Bigl (1+\frac{1}{z}\, 
\p_{\beta}\p_{t_{\alpha , 1}}F
\Bigr ) +O(z^{-2}),
$$ 
i.e.,
$
c_1^{(\alpha )}S'(\eta_{\alpha \beta})= \p_{\beta}\p_{t_{\alpha , 1}}F.
$
Using relations (\ref{E1b}), (\ref{E1c}), (\ref{S'}), we arrive at
the equation
\beq\label{E1e}
e^{(\p_{\beta}-\p_{\alpha})\p_{\alpha}F}
\p_{\beta}\p_{t_{\alpha , 1}}F=
\epsilon_{\beta \alpha} \frac{\theta_4^2(0)\, \theta_2(\eta_{\alpha \beta})
\, \theta_3(\eta_{\alpha \beta})}{\theta_2(0)\, 
\theta_3(0)\,\theta_4^2(\eta_{\alpha \beta})}
\eeq
which we will need later.

Remarkably, the substitution (\ref{E1}) solves the 4-point relations
(\ref{40c}) and (\ref{31c}) converting them into identities. Indeed,
substituting (\ref{E1}) into 4-point relation (\ref{40c}), we see that
it is satisfied identically by virtue of the identity
\beq\label{40d}
\begin{array}{l} 
\displaystyle{
\frac{\theta_1(u_{\beta}-u_{\nu})\, 
\theta_1(u_{\beta}\! -\! u_{\mu})\, 
\theta_1(u_{\mu}\! -\! u_{\nu})}{\theta_4(u_{\beta}\! -\! u_{\nu})\, 
\theta_4(u_{\beta}\! -\! u_{\mu})\, 
\theta_4(u_{\mu}\! -\! u_{\nu})}-
\frac{\theta_1(u_{\alpha}\! -\! u_{\nu})\, 
\theta_1(u_{\alpha}\! -\! u_{\mu})\, 
\theta_1(u_{\mu}\! -\! u_{\nu})}{\theta_4(u_{\alpha}\! -\! u_{\nu})\, 
\theta_4(u_{\alpha}\! -\! u_{\mu})\, 
\theta_4(u_{\mu}\! -\! u_{\nu})}}
\\ \\
\displaystyle{
+\, \frac{\theta_1(u_{\alpha}-u_{\beta})\, 
\theta_1(u_{\alpha}\! -\! u_{\nu})\, 
\theta_1(u_{\beta}\! -\! u_{\nu})}{\theta_4(u_{\alpha}\! -\! u_{\beta})\, 
\theta_4(u_{\alpha}\! -\! u_{\nu})\, 
\theta_4(u_{\beta}\! -\! u_{\nu})}-
\frac{\theta_1(u_{\alpha}\! -\! u_{\beta})\, 
\theta_1(u_{\alpha}\! -\! u_{\mu})\, 
\theta_1(u_{\beta}\! -\! u_{\mu})}{\theta_4(u_{\alpha}\! -\! u_{\beta})\, 
\theta_4(u_{\alpha}\! -\! u_{\mu})\, 
\theta_4(u_{\beta}\! -\! u_{\mu})}=0,}
\end{array}
\eeq
where we do not indicate the dependence of all the variables on 
$z$. The proof of this identity
is standard. It consists in three steps. First, one should make sure that the
left-hand side is an elliptic (i.t., double-periodic) function of one 
of the variables (say, $u_{\alpha}$). This can be done using the monodromy 
properties of the theta-functions listed in the appendix. Next, one should
check that residues at the simple poles coming from zeros of the 
denominators in (\ref{40d}) vanish. Therefore, the left-hand side is 
a constant independent of $u_{\alpha}$ (but in principle could depend on
the other variables). Finally, the constant can be found by putting
$u_{\alpha}$ equal to, say, $u_{\nu}$, then the left-hand side of
(\ref{40d}) is zero (two terms vanish, and the other two cancel each
other). 
In its turn, equation (\ref{31c}) after the substitution 
(\ref{E1}) is satisfied identically, too\footnote{This follows from
the same identity (\ref{40d}) in which one of the $u$-variables is shifted
by $\tau/2$, then $\mbox{sn}(u+\frac{\tau}{2})=\mbox{sn}^{-1}(u)$.}.

Moreover, one can see that the same substitution converts into
identities {\it all} $P$-point 
dispersionless Hirota-Miwa equations of the general form (\ref{genHM2}).
Indeed, let us rewrite (\ref{genHM2}) in the form
\beq\label{genHM3}
\begin{array}{l}
\displaystyle{
\sum_{s=1}^{P^+}
\prod_{{\scriptsize \begin{array}{l}i=1\\ i\neq s \end{array}}}^{P^+}
\left (E_{\alpha_i \alpha_s}(a_s, a_i)
e^{\nabla_{\alpha_s}(a_s)\nabla_{\alpha_i}(a_i)}\right )^{-1}
\prod_{k=1}^{P^-}
\left (E_{\beta_k \alpha_s}(a_s, b_k)
e^{\nabla_{\alpha_s}(a_s)\nabla_{\beta_k}(b_k)}\right )
}
\\ \\
\displaystyle{
+\sum_{s=1}^{P^-}
\prod_{{\scriptsize \begin{array}{l}k=1\\ k\neq s \end{array}}}^{P^-}
\left (E_{\beta_k \beta_s}(b_s, b_k)
e^{\nabla_{\beta_s}(b_s)\nabla_{\beta_k}(b_k)}\right )^{-1}
\prod_{i=1}^{P^-}
\left (E_{\alpha_i \beta_s}(b_s, a_i)
e^{\nabla_{\beta_s}(b_s)\nabla_{\alpha_i}(a_i)}\right )
}
=0.
\end{array}
\eeq
which is most convenient to make the substitution (\ref{E1}).
Denote for brevity
$$
u_i=u_{\alpha_i}(a_i), \qquad v_k=u_{\beta_k}(b_k), \quad i,k=1, \ldots , M,
$$
then (\ref{genHM3}) converts into
\beq\label{id1}
\begin{array}{l}
\displaystyle{
\sum_{s=1}^{P^+}
\prod_{{\scriptsize \begin{array}{l}i=1\\ i \neq s \end{array}}}^{P^+}
\frac{\theta_4 (u_i-u_s)}{\theta_1 (u_i-u_s)}\,
\prod_{k=1}^{P^-}\frac{\theta_1 (u_s-v_k)}{\theta_4 (u_s-v_k)}}
\\ \\
\displaystyle{\phantom{aaaaaaaaaaaaaaa}
 +\, \sum_{s=1}^{P^-}
\prod_{{\scriptsize \begin{array}{l}m=1\\ m\neq s \end{array}}}^{P^-}
\frac{\theta_4 (v_m -v_s)}{\theta_1 (v_m -v_s)}\,
\prod_{l=1}^{P^-}\frac{\theta_1 (v_s-u_l)}{\theta_4 (v_s -u_l)}=0.}
\end{array}
\eeq
Shifting the $v$-variables as $v_i \to v_i +\frac{\tau}{2}$,
we rewrite this in the following equivalent form:
\beq\label{id2}
\begin{array}{l}
\displaystyle{
\sum_{s=1}^{P^+}
\prod_{{\scriptsize \begin{array}{l}i=1\\ i \neq s \end{array}}}^{P^+}
\frac{\theta_4 (u_i-u_s)}{\theta_1 (u_i-u_s)}\,
\prod_{k=1}^{P^-}\frac{\theta_4 (u_s-v_k)}{\theta_1 (u_s-v_k)}}
\\ \\
\displaystyle{\phantom{aaaaaaaaaaaaaaa}
 +\, \sum_{s=1}^{P^-}
\prod_{{\scriptsize \begin{array}{l}m=1\\ m\neq s \end{array}}}^{P^-}
\frac{\theta_4 (v_m -v_s)}{\theta_1 (v_m -v_s)}\,
\prod_{l=1}^{P^-}\frac{\theta_4 (v_s-u_l)}{\theta_1 (v_s -u_l)}=0.}
\end{array}
\eeq
To see that this holds identically for all variables
$\{u_i\}$, $\{v_k\}$, it is enough to notice that the left-hand 
side is proportional to the sum of residues of the following
elliptic function with periods $1, \tau$:
\beq\label{id4}
f(u)=\prod_{i=1}^{P^+} \frac{\theta_4(u-u_i)}{\theta_1(u-u_i)}
\prod_{j=1}^{P^-} \frac{\theta_4(u-v_j)}{\theta_1(u-v_j)},
\eeq
which is zero. (Note that the condition that $P^++P^-\in 2\ZZ_{+}$ is 
important for this.)
Therefore, the substitution (\ref{E1}) solves all
equations of the hierarchy simultaneously, and thus the whole hierarchy
is actually equivalent to it.

Note that all this holds for any elliptic modular parameter $\tau$,
including the degenerate cases $\tau \to +i0$ or $\tau \to +i\infty$, in which
elliptic functions become trigonometric or hyperbolic.
However, as is shown below, consistency of the whole construction
requires a fixation of $\tau$ in some particular way, so that it becomes
a dynamical variable depending on the times ${\bf t}$.

\subsection{The elliptic curve}

To determine the modular parameter,
we should make explicit the elliptic curve hidden in the 
hierarchy. This can be done by considering degenerate cases of the
4-point relations. To wit, we put $\alpha =\beta =\nu \neq \mu$ and
tend $c,d\to \infty$ in (\ref{40c}), (\ref{31c}). Renaming 
$\mu \leftrightarrow \beta$ after this, we see that equation (\ref{40c})
yields
\beq\label{E2}
\begin{array}{l}
-\, b^{-1}e^{(\nabla_{\alpha}(b)\p_{\alpha}+\nabla_{\alpha}(b)
\p_{\beta}+\p_{\alpha}\p_{\beta})F}
+\, a^{-1}e^{(\nabla_{\alpha}(a)\p_{\alpha}+\nabla_{\alpha}(a) \p_{\beta}
+\p_{\alpha}\p_{\beta})F}
\\ \\
\phantom{aaaaaaaaa}
-\, (a^{-1}-b^{-1})e^{(\nabla_{\alpha}(a)\nabla_{\alpha}(b)
+\nabla_{\alpha}(a) \p_{\beta}+\nabla_{\alpha}(b) \p_{\beta})F}
\\ \\
\phantom{aaaaaaaaaaaaaaaa}
+\, (a^{-1}-b^{-1})(ab)^{-1}e^{(\nabla_{\alpha}(a)\nabla_{\alpha}(b)
+\nabla_{\alpha}(a) \p_{\alpha}+\nabla_{\alpha}(b) \p_{\alpha})F}=0.
\end{array}
\eeq
In a similar way, equation (\ref{31c}) yields:
\beq\label{E3}
\begin{array}{l}
b^{-1}e^{(\nabla_{\alpha}(b)\p_{\alpha}+\nabla_{\alpha}(a)
\p_{\beta})F}
- a^{-1}e^{(\nabla_{\alpha}(a)\p_{\alpha}+\nabla_{\alpha}(b) \p_{\beta})F}
+(a^{-1}-b^{-1})e^{(\nabla_{\alpha}(a)\nabla_{\alpha}(b)+\p_{\alpha}
\p_{\beta})F}
\\ \\
\phantom{aaaa}
-\, (a^{-1}-b^{-1})(ab)^{-1}e^{(\nabla_{\alpha}(a)\nabla_{\alpha}(b)
+\nabla_{\alpha}(a)\p_{\alpha}+\nabla_{\alpha}(a) \p_{\beta}
+\nabla_{\alpha}(b)\p_{\alpha} +\nabla_{\alpha}(b)\p_{\beta}
+\p_{\alpha}\p_{\beta})F}=0.
\end{array}
\eeq
These two equation have to be satisfied simultaneously, and this requirement 
allows one to recover the elliptic curve and make it explicit, i.e., 
to represent its points as solutions of a polynomial equation in two complex
variables.

To represent the equations in a more suggestive form,
introduce the following functions\footnote{The same notation 
from \cite{SZ24} corresponds in fact to 
$1/w_{\alpha}$, $1/w_{\alpha \beta}$ for the $w$'s introduced
here;
note, however, that this does not matter because the equations
are invariant under this transformation.}:
\beq\label{E4}
w_{\alpha}(z)=z^{-1}e^{\nabla_{\alpha}(z)\p_{\alpha}F}, \quad
w_{\alpha \beta}(z)=e^{\nabla_{\alpha}(z)\p_{\beta}F} \;\; (\mbox{for}\; 
\alpha \neq \beta ).
\eeq

In this notation, equations (\ref{E2}), (\ref{E3}) acquire the form

\beq\label{E6}
\begin{array}{l}
R_{\alpha \beta}\Bigl (w_{\alpha}(a)w_{\alpha \beta}(a)-
w_{\alpha}(b)w_{\alpha \beta}(b)\Bigr )
\\ \\
\phantom{aaaaaaaaaaa}
=(a^{-1}-b^{-1})e^{\nabla_{\alpha}(a)\nabla_{\alpha}(b)F}
\Bigl (w_{\alpha \beta}(a)w_{\alpha \beta}(b)-
w_{\alpha}(a)w_{\alpha}(b)\Bigr ),
\end{array}
\eeq

\beq\label{E7}
\begin{array}{l}
\phantom{aa}w_{\alpha}(a)w_{\alpha \beta}(b)-
w_{\alpha}(b)w_{\alpha \beta}(a)
\\ \\
\phantom{aaaaaaaaaaa}
=R_{\alpha \beta}(a^{-1}-b^{-1})e^{\nabla_{\alpha}(a)\nabla_{\alpha}(b)F}
\Bigl (1- w_{\alpha}(a)w_{\alpha}(b)w_{\alpha \beta}(a)w_{\alpha \beta}(b)
\Bigr ).
\end{array}
\eeq

\noindent
Dividing the left- and right-hand sides of the first equation by 
the corresponding sides of the second one, 
we see that $e^{\nabla_{\alpha}(a)\nabla_{\alpha}(b)F}$ cancels, and the
result can be written entirely in terms of the functions $w_{\alpha}$ and
$w_{\alpha \beta}$. After some simple
transformations it can be represented in the form
$$
\begin{array}{l}
\displaystyle{
\phantom{aa}R_{\alpha \beta}^2 \left (w_{\alpha}(a)w_{\alpha \beta}(a)+
\frac{1}{w_{\alpha}(a)w_{\alpha \beta}(a)}\right ) -
\frac{w_{\alpha}(a)}{w_{\alpha \beta}(a)}-
\frac{w_{\alpha \beta}(a)}{w_{\alpha}(a)}}
\\ \\
=\, \displaystyle{
R_{\alpha \beta}^2 \left (w_{\alpha}(b)w_{\alpha \beta}(b)+
\frac{1}{w_{\alpha}(b)w_{\alpha \beta}(b)}\right ) -
\frac{w_{\alpha}(b)}{w_{\alpha \beta}(b)}-
\frac{w_{\alpha \beta}(b)}{w_{\alpha}(b)}}.
\end{array}
$$
The left-hand side depends only on $a$ while the right-hand side depends
only on $b$, whence both equal to some constant\footnote{Meaning that
it does not depend on $a$ or $b$, but, of course, it can and does depend
on the times.} which we denote
as $-V_{\alpha \beta}$:
\beq\label{E8}
R_{\alpha \beta}^2 \left (w_{\alpha}(z)w_{\alpha \beta}(z)+
\frac{1}{w_{\alpha}(z)w_{\alpha \beta}(z)}\right ) -
\frac{w_{\alpha}(z)}{w_{\alpha \beta}(z)}-
\frac{w_{\alpha \beta}(z)}{w_{\alpha}(z)}=-V_{\alpha \beta}, \quad
\alpha \neq \beta .
\eeq
The constant can be found from the $z\to \infty$ limit of this equation.
Expanding the equations (\ref{E4}) in powers of $z$ as $z\to \infty$
and substituting into (\ref{E8}),
we obtain, comparing the leading terms:
\beq\label{V}
V_{\alpha \beta}=2e^{(\p_{\beta}-\p_{\alpha})\p_{\alpha}F}
\p_{\beta}\p_{t_{\alpha , 1}}F.
\eeq
From (\ref{E1e}) we conclude that
\beq\label{V1}
V_{\alpha \beta}=
2\epsilon_{\beta \alpha} \frac{\theta_4^2(0)\, \theta_2(\eta_{\alpha \beta})
\, \theta_3(\eta_{\alpha \beta})}{\theta_2(0)\, 
\theta_3(0)\,\theta_4^2(\eta_{\alpha \beta})}.
\eeq
Note that the right-hand side implies that 
$V_{\beta \alpha}=-V_{\alpha \beta}$, although this can not be seen from
the original definition (\ref{V}).

Equation (\ref{E8}) can be written in the
polynomial form:
\beq\label{E9}
P(w_{\alpha}, w_{\alpha \beta})=
R_{\alpha \beta}^2 \Bigl (w_{\alpha}^2 w_{\alpha \beta}^2 +1\Bigr )-
\Bigl (w_{\alpha}^2 + w_{\alpha \beta}^2\Bigr )+V_{\alpha \beta}
w_{\alpha}w_{\alpha \beta}=0.
\eeq
The polynomial $P(w_{\alpha}, w_{\alpha \beta})$ is quadratic in each of
the two complex variables, whence the equation 
$P(w_{\alpha}, w_{\alpha \beta})=0$ 
defines an elliptic curve. The functions (\ref{E4}) are meromorphic 
functions on this curve, and $z^{-1}$ plays the role of a local parameter
in a neighborhood of $\infty$. Both functions are regular at $\infty$,
and $w_{\alpha}$ has there a simple zero.

It is well known that any algebraic curve defined by the equation $P(x,y)=0$
with a bi-quadratic polynomial $P(x,y)$ can be uniformized by
elliptic functions
The right-hand side of equation (\ref{E1a}) just provides such
uniformization for the curve (\ref{E9}) in the explicit form:
\beq\label{uni}
w_{\alpha}(z)=
\frac{\theta_1(u_{\alpha}(z)-\eta_{\alpha})}{\theta_4(u_{\alpha}(z)-
\eta_{\alpha})}, \qquad
w_{\alpha \beta}(z)=\epsilon_{\beta \alpha}
\frac{\theta_1(u_{\alpha}(z)-\eta_{\beta})}{\theta_4(u_{\alpha}(z)-
\eta_{\beta})}.
\eeq 
This means that the equation of the curve is satisfied identically if one
substitutes into (\ref{E9}) 
$w_{\alpha}$, $w_{\alpha \beta}$, 
$R_{\alpha \beta}$, $V_{\alpha \beta}$ as they are 
expressed in (\ref{uni}),
(\ref{E5}), (\ref{V1}) respectively. Some details on uniformization
of elliptic curves by theta-functions not mentioned here 
can be found in Appendix D to the 
paper \cite{SZ24}.

Now we are ready to answer the question how the
modular parameter $\tau$ should be chosen. 
Note first of all that
\beq\label{E10}
\frac{V_{\alpha \beta}}{R_{\alpha \beta}}=
\frac{2\, S'(\eta_{\alpha \beta})}{\pi \theta_2(0)\, \theta_3(0)}
\eeq
(see (\ref{E1b}), (\ref{S'}) and (\ref{V1})).
Then identity (\ref{S1}) can be written as
$$
R_{\alpha \beta}^2 +R_{\alpha \beta}^{-2} -
\Bigl ( \frac{V_{\alpha \beta}}{2R_{\alpha \beta}}\Bigr )^2 =
\frac{\theta_2^2 (0)}{\theta_3^2 (0)}+\frac{\theta_3^2 (0)}{\theta_2^2 (0)},
$$
or
\beq\label{E11}
e^{2\p_{\alpha}\p_{\beta}F}+e^{-2\p_{\alpha}\p_{\beta}F}-
e^{-2\p_{\alpha}^2F}(\p_{\beta}\p_{t_{\alpha ,1}}F)^2 =
\frac{\theta_2^2 (0|\tau )}{\theta_3^2 (0|\tau )}+
\frac{\theta_3^2 (0|\tau )}{\theta_2^2 (0|\tau )},
\eeq
where we explicitly indicate the dependence of the theta-functions
on the modular parameter $\tau$. 
Being written in this form, (\ref{E11}) is the key equation which allows one to
fix the modular parameter $\tau$ consistently. Indeed, the right-hand side 
depends only on $\tau$ while the left-hand side in general has a 
non-trivial dependence on all the times ${\bf t}$ determined by 
equations of the hierarchy, 
so $\tau =\tau ({\bf t})$ is a dynamical variable.

Note that the right-hand side of (\ref{E11}) 
does not contain the indices $\alpha$, $\beta$,
and, in particular, is symmetric under their permutation. For the
left-hand side this implies
$$
e^{-2\p_{\alpha}^2F}(\p_{\beta}\p_{t_{\alpha ,1}}F)^2=
e^{-2\p_{\beta}^2F}(\p_{\alpha}\p_{t_{\beta ,1}}F)^2,
$$
which does hold by virtue of $V_{\alpha \beta}=-V_{\beta \alpha}$,
or
\beq\label{E12}
R_{\alpha}(\p_{\beta}\p_{t_{\alpha ,1}}F)=-
R_{\beta}(\p_{\alpha}\p_{t_{\beta ,1}}F).
\eeq

\subsection{Another form of the elliptic curve}

For completeness, let us mention that the same elliptic curve can be
equivalently defined by other polynomial equations as well. One of them
was given in \cite{SZ24}. In addition to (\ref{E4}), introduce the 
functions
\beq\label{p0}
p_{\alpha}(z)=z-\nabla_{\alpha}(z)\p_{t_{\alpha , 1}}F,
\qquad
p_{\alpha \beta}(z)=-\nabla_{\alpha}(z)\p_{t_{\beta, 1}}F.
\eeq
In \cite{SZ24} it was shown that they satisfy the polynomial equations
\beq\label{p1}
\begin{array}{l}
R_{\alpha}^2 \Bigl (w_{\alpha}^2(z)+w_{\alpha}^{-2}(z)\Bigr )=p_{\alpha}^2(z)
+V_{\alpha},
\\ \\
R_{\alpha}^2 \Bigl (w_{\beta \alpha}^2(z)+w_{\beta 
\alpha}^{-2}(z)\Bigr )=p_{\beta \alpha}^2(z)
+V_{\alpha} \quad \mbox{for $\alpha \neq \beta$},
\end{array}
\eeq
where
\beq\label{p2}
V_{\alpha}=(\p_{\alpha}\p_{t_{\alpha , 1}}F)^2 +2 \p_{t_{\alpha ,1}}^2 F-
\p_{\alpha}\p_{t_{\alpha , 2}}F.
\eeq
This follows from some degenerate 2-point Hirota-Miwa equations 
containing derivatives with respect to $t_{\alpha , 1}$.
Equations (\ref{p1}) define the same curve, with $p_{\alpha}, p_{\beta \alpha}$
being some other algebraic functions 
on it\footnote{In this paper we do not consider the functions (\ref{p0}) 
and curves of the form (\ref{p1}) in detail because this requires 
an analysis of Hirota-Miwa equations containing derivatives with
respect to times $t_{\alpha , k}$ with $k\geq 1$. We hope to present
some details on
this point elsewhere.}. 
Their uniformization through
the theta-functions is as follows:
\beq\label{p3}
\begin{array}{l}
\displaystyle{
p_{\alpha}(z)=\frac{\gamma_{\alpha}}{\pi}S'(u_{\alpha}(z)-\eta_{\alpha}),
\qquad
p_{\beta \alpha}(z)=\frac{\gamma_{\alpha}}{\pi}
S'(u_{\beta}(z)-\eta_{\alpha})},
\end{array}
\eeq
where $\gamma_{\alpha}=\gamma_{\alpha}({\bf t})$ is a dynamical variable and
the elliptic function $S'(u)$ is given by (\ref{S'}). These formulas 
should be supplemented by expressions for $R_{\alpha}$ and $V_{\alpha}$:
\beq\label{p4}
R_{\alpha}=\gamma_{\alpha}\theta_2(0)\theta_3(0), \qquad
V_{\alpha}=\gamma_{\alpha}^2 \Bigl (\theta_2^4(0)+\theta_3^4(0)\Bigr ).
\eeq
Substituting all this into equations (\ref{p1}), one can see that they 
are satisfied identically.

Equations (\ref{p4}) give an alternative way to determine the modular
parameter. From them we conclude that
\beq\label{E11a}
\frac{V_{\alpha}}{R^2_{\alpha}}
 =e^{-2\p_{\alpha}^2F}\Bigl ((\p_{\alpha}
 \p_{t_{\alpha , 1}}F)^2 +2 \p_{t_{\alpha ,1}}^2 F-
\p_{\alpha}\p_{t_{\alpha , 2}}F\Bigr )=
\frac{\theta_2^2 (0|\tau )}{\theta_3^2 (0|\tau )}+
\frac{\theta_3^2 (0|\tau )}{\theta_2^2 (0|\tau )},
\eeq
where the right-hand side (the same as in (\ref{E11})) depends only
on the modular parameter $\tau$ while the left-hand side in general 
depends on all the times according to equations of the hierarchy.

\subsection{Elliptic uniformization of the multi-component Pfaff-Toda
hierarchy}
\label{section:ellipticToda}

According to the Pfaff-Toda-DKP equivalence 
established in section \ref{section:TodaDKP}
(see, in particular, (\ref{re}) and (\ref{tautau})),
it remains to rewrite equation (\ref{E1}) in terms of the variables specific to
the Pfaff-Toda hierarchy.

So, we introduce the bar-counterpart of the vector field (\ref{n1}):
\beq\label{n1a}
\bar \nabla_{\alpha}(z)=\bar \p_{\alpha}+\sum_{k\geq 1}\frac{z^{-k}}{k}
\p_{\bar t_{\alpha , 1}},
\eeq
where
$
\bar \p_{\alpha}\equiv \p_{\bar t_{\alpha , 0}}.
$
Also, we need to take into account that the $F$-functions of the two
hierarchies slightly differ because of the sign factor in (\ref{tautau}).
Writing it as
$$
(-1)^{\frac{1}{2}|\bar {\bf n}|(|\bar {\bf n}|-1)}=
e^{\frac{i\pi}{2}|\bar {\bf n}|(|\bar {\bf n}|-1)},
$$
we have:
$$
\frac{i\pi}{2}|\bar {\bf n}|(|\bar {\bf n}|-1)=
\frac{i\pi}{2}\left (\sum_{\mu , \nu}\bar n_{\mu}\bar n_{\nu}-
\sum_{\mu}\bar n_{\mu}\right )=
\frac{i\pi}{2\hbar^2}\left (\sum_{\mu , \nu}\bar t_{\mu ,0}\bar t_{\nu ,0}-
\hbar \sum_{\mu}\bar t_{\mu ,0}\right ),
$$
from which one can see that
the relation between the two $F$-functions
is as follows:
\beq\label{FF}
F^{\rm Toda}({\bf t}, \bar {\bf t})
= \frac{i\pi}{2}\sum_{\mu , \nu}\bar t_{\mu ,0}\bar t_{\nu ,0} +
F^{\rm DKP}(\tilde {\bf t}),
\eeq
where the sets of times are
$$
{\bf t}=\{{\bf t}_1, \ldots , {\bf t}_{N}\}, 
\qquad
\bar {\bf t}=\{ \bar {\bf t}_1, \ldots , 
\bar {\bf t}_{N}\}, \qquad
\tilde {\bf t}=\{{\bf t}_1, \ldots , {\bf t}_{2N}\},
$$
with ${\bf t}_{\mu}=\tilde {\bf t}_{\mu}$,
$\bar {\bf t}_{\mu}=\tilde {\bf t}_{\mu +N}$ for $\mu =1, \ldots , N$,
and
$$
{\bf t}_{\gamma}=\{ t_{\gamma , 0}, t_{\gamma , 1}, \ldots , \}, 
\qquad
\bar {\bf t}_{\gamma}=\{ \bar t_{\gamma , 0}, \bar t_{\gamma , 1}, \ldots , \}.
$$
Therefore, translating equation (\ref{E1}) to the Toda variables, we should
take into account that 
$$
e^{\nabla_{\alpha +N}(a)\nabla_{\beta +N}(b)F^{\rm DKP}} \to -\, 
e^{\bar \nabla_{\alpha}(a)\bar \nabla_{\beta}(b)F^{\rm Toda}},
\quad \alpha , \beta =1, \ldots , N.
$$
This can be done by introducing, along with the $u_{\alpha}$'s functions
$\bar u_{\alpha}(z)=-u_{\alpha +N}(z)$. Their expansion around $\infty$
is similar to (\ref{u}):
\beq\label{baru}
\bar u_{\alpha}(z)=\bar \eta_{\alpha}({\bf t}, \bar {\bf t})
+\sum_{k\geq 1}\bar c_k^{(\alpha )}({\bf t}, \bar {\bf t}) z^{-k}.
\eeq
The final result is the following system of equations:
\beq\label{final}
\left \{
\begin{array}{rcl}
E_{\beta \alpha}(a,b) e^{\nabla_{\alpha}(a)\nabla_{\beta}(b)F}
&=& \displaystyle{\frac{\theta_1(u_{\alpha}(a)- 
u_{\beta}(b))}{\theta_4(u_{\alpha}(a)-
u_{\beta}(b))},}
\\ && \\
e^{\nabla_{\alpha}(a)\bar \nabla_{\beta}(b)F}& =& \displaystyle{
\frac{\theta_1(u_{\alpha}(a)+ \bar u_{\beta}(b))}{\theta_4(u_{\alpha}(a)+
\bar u_{\beta}(b))},}
\\ && \\
E_{\beta \alpha}(a,b) e^{\bar \nabla_{\alpha}(a)\bar \nabla_{\beta}(b)F}
& =& \displaystyle{\frac{\theta_1(\bar u_{\alpha}(a)- 
\bar u_{\beta}(b))}{\theta_4(\bar u_{\alpha}(a)-
\bar u_{\beta}(b))}.}
\end{array} \right.
\eeq
Here $\alpha , \beta =1, \ldots , N$.
Our definition of the $\bar u_{\alpha}$'s is such that the bar-counterpart
of equations (\ref{E1}) has exactly the same form, with 
$\bar u_{\alpha}$-functions
instead of $u_{\alpha}$'s. However, because of this
the equation
that mixes the variables with and without bar (the second line in (\ref{final}))
contains $u_{\alpha}+\bar u_{\beta}$ rather than
$u_{\alpha}-\bar u_{\beta}$. At $N=1$ equations (\ref{final})
coincide with the result obtained earlier in \cite{AZ15} for the 
one-component dispersionless Pfaff-Toda hierarchy.

At last let us comment on the choice of the sign factors in
(\ref{final}). On the DKP side, the $2N$ discrete variables are 
ordered as $\{n_1, \ldots , n_N, \bar n_1 , \ldots , \bar n_N\}$, and
the $\epsilon$-symbols are defined according to this order, and that is
why there are $\epsilon_{\beta \alpha}$ in the first and the third
equation in (\ref{final}). However, from the DKP point of view,
the index $\alpha$ in the second equation
should be regarded as being always less than $\beta$
(even if $\alpha \geq \beta$), i.e., strictly speaking, the
right-hand side should go with sign ``$-$''. But, according to
(\ref{trans}) the $F$-function is defined up to the term 
$\displaystyle{i\pi \sum_{\mu ,\nu}t_{\mu ,0}\bar t_{\nu , 0} }$, 
which contributes as
$$
\exp \Bigl (i\pi\p_{\alpha}\bar \p_{\beta}
\sum_{\mu ,\nu}t_{\mu ,0}\bar t_{\nu , 0}\Bigr ) =e^{i\pi}=-1,
$$
and
this freedom just allows one to choose the ``$+$'' sign, as in
(\ref{final}).

\section{Conclusion and discussion}

The main results of this paper are:
\begin{itemize}
\item
The equivalence of the $N$-component Pfaff-Toda and the 
$2N$-component DKP hierarchies is established;
\item
Using this equivalence, the dispersionless version of the multi-component
Pfaff-Toda hierarchy in the elliptic parametrization is given explicitly
(equations (\ref{final})).
\end{itemize}

Like in the DKP case, there is an elliptic curve underlying the
dispersionless version of the hierarchy.
It is worth noting, that although the modular parameter $\tau$ of the
elliptic functions used for uniformization of the elliptic curve is a dynamical
variable, the formal limit $\tau \to +i\infty$ converts equation (\ref{E1}) 
into the one that emerges in the dispersionless multi-component KP hierarchy
(see \cite{Z24}). In this case the curve becomes rational and 
the elliptic functions degenerate to trigonometric ones. So, in some sense
the Pfaff-type hierarchies can be regarded as elliptic deformations of
usual KP and Toda hierarchies. 

A closely related problem is as follows. It is known that the 
one-component DKP hierarchy
is a ``half'' of the so-called large BKP hierarchy (see, e.g.,
\cite{book}), which was recently rediscovered 
in \cite{KZ22,PZ23} as the Toda hierarchy with constraint of type B.
More precisely, the DKP equations connect tau-functions
$\tau (n, {\bf t})$ in which the discrete variable $n\in \ZZ$ has a definite
parity, so there are two disconnected ``sectors'': even (for even $n\in 2\ZZ$)
and odd (for odd $n\in 2\ZZ +1$). Therefore, there are two independent 
DKP hierarchies, ``even'' and ``odd''. The full large BKP hierarchy 
contains more equations. These additional equations
mix the two sectors, i.e., they connect
the tau-functions $\tau (n, {\bf t})$ with $n$'s of different parities.
As it was shown in \cite{Z24a}, in the dispersionless limit the additional
equations make the underlying elliptic curve degenerate (i.e., rational).
However, in contrast to the KP case mentioned above, this degeneration
corresponds to the limit $\tau \to +i0$ rather than $\tau \to +i\infty$. 
For multi-component hierarchies, there are two disconnected copies
of the DKP hierarchy,
too, one acting in the even sector ($|{\bf n}|\in 2\ZZ$) and another in 
the odd one ($|{\bf n}|\in 2\ZZ +1$). As far as we know, the hypothetical
multi-component large BKP hierarchy, at least in the 
bilinear formalism, was not discussed in the literature. An interesting
problem for future research is to 

An important still unsolved problem is to figure out what is a minimal set
of the Hirota-Miwa equations that are equivalent to the whole hierarchy.
For one-component KP, modified KP and BKP hierarchies 
the answer is known \cite{Shigyo13}: just the very first nontrivial
Hirota-Miwa relation (containing three or four terms)
is already equivalent to the whole hierarchy. 
To the best of our knowledge, nothing in this respect is known for 
multi-component hierarchies and for hierarchies of the Pfaff type,
even in the one-component case. Our hypothesis is that in the Pfaff case
the two non-degenerate 4-point 
relations (\ref{40a}), (\ref{31a}) 
are equivalent to the whole hierarchy, but this is probably not the minimal set,
and we can not exclude that the 4-point 
relations are corollaries of some of the 2-point ones listed in
the appendix of the paper \cite{SZ25a}.

\section*{Appendix: Theta-functions}

\addcontentsline{toc}{section}{Appendix: Theta-functions}
\def\theequation{A\arabic{equation}}
\def\theHequation{\theequation}
\setcounter{equation}{0}

The Jacobi's theta-functions $\theta_a (u)=
\theta_a (u|\tau )$, $a=1,2,3,4$, are defined by the absolutely 
convergent infinite sums as follows:
\beq\label{Bp1}
\begin{array}{l}
\theta _1(u)=-\displaystyle{\sum _{k\in \z}}
\exp \left (
\pi i \tau (k+\frac{1}{2})^2 +2\pi i
(u+\frac{1}{2})(k+\frac{1}{2})\right ),
\\ \\
\theta _2(u)=\displaystyle{\sum _{k\in \z}}
\exp \left (
\pi i \tau (k+\frac{1}{2})^2 +2\pi i
u(k+\frac{1}{2})\right ),
\\ \\
\theta _3(u)=\displaystyle{\sum _{k\in \z}}
\exp \left (
\pi i \tau k^2 +2\pi i u k \right ),
\\ \\
\theta _4(u)=\displaystyle{\sum _{k\in \z}}
\exp \left (
\pi i \tau k^2 +2\pi i
(u+\frac{1}{2})k\right ),
\end{array}
\eeq where $\tau$ is a complex parameter (the modular parameter) 
such that ${\rm Im}\, \tau >0$. The function 
$\theta_1(u)$ is odd, the other three functions are even.
The infinite product representation for the $\theta_1(u)$ reads: 
\beq
\label{infprod}
\theta_1(u|\tau)=2q^{\frac{1}{4}}\sin\pi u
\prod_{n=1}^\infty(1-q^{2n})(1-q^{2n}e^{2\pi i u})(1-q^{2n}e^{-2\pi i u}),
\eeq 
where $q=e^{\pi i \tau}$.
We also mention the identity
\beq\label{theta1prime}
\theta_1'(0)=\pi \theta_2(0) \theta_3(0) \theta_4(0).
\eeq

Here we list the transformation properties of the theta functions.

\smallskip

\noindent
Shifts by periods:
\beq\label{p}
\begin{array}{l}
\theta_1(u+1)=-\theta_1(u),\\ \\
\theta_2(u+1)=-\theta_2(u),\\ \\
\theta_3(u+1)=\theta_3(u),\\ \\
\theta_4(u+1)=\theta_4(u).
\end{array}\hspace{2cm}
\begin{array}{l}
\theta_1(u+\tau)=-e^{-\pi i(2u+\tau)}\theta_1(u),\\ \\
\theta_2(u+\tau)=e^{-\pi i(2u+\tau)}\theta_2(u),\\ \\
\theta_3(u+\tau)=e^{-\pi i(2u+\tau)}\theta_3(u),\\ \\
\theta_4(u+\tau)=-e^{-\pi i(2u+\tau)}\theta_4(u).
\end{array}\vspace{0.3cm}
\eeq
Shifts by half-periods:
\beq\label{hp}
\begin{array}{l}
\theta_1(u+{\textstyle\frac{1}{2}})=\theta_2(u),\\ \\
\theta_2(u+{\textstyle\frac{1}{2}})=-\theta_1(u),\\ \\
\theta_3(u+{\textstyle\frac{1}{2}})=\theta_4(u),\\ \\
\theta_4(u+{\textstyle\frac{1}{2}})=\theta_3(u).
\end{array}\hspace{2cm}
\begin{array}{l}
\theta_1(u+{\textstyle\frac{\tau}{2}})=
ie^{-\pi i(u+\tau/4)}\theta_4(u),\\ \\
\theta_2(u+{\textstyle\frac{\tau}{2}})=
e^{-\pi i(u+\tau/4)}\theta_3(u),\\ \\
\theta_3(u+{\textstyle\frac{\tau}{2}})=
e^{-\pi i(u+\tau/4)}\theta_2(u),\\ \\
\theta_4(u+{\textstyle\frac{\tau}{2}})=i
e^{-\pi i(u+\tau/4)}\theta_1(u).
\end{array}\vspace{0.3cm}
\eeq

Let us also mention here the identity which is used in section
\ref{section:elliptic}:
\beq\label{id}
\theta_2^4(0)\, 
\frac{\theta_2^2 (u)\, \theta_3^2 (u)}{\theta_1^2 (u)\, \theta_4^2 (u)}=
\theta_2^2 (0)\, \theta_3^2 (0)\left (
\frac{\theta_4^2(u)}{\theta_1^2(u)}+ \frac{\theta_1^2(u)}{\theta_4^2(u)}
\right ) - \Bigl (\theta_2^4(0) +\theta_3^4(0)\Bigr ).
\eeq

For a more detailed account of properties of the theta-functions
see \cite{KZ15}.

\section*{Acknowledgments}
\addcontentsline{toc}{section}{Acknowledgments}

This work is an output of the research project 
``Symmetry. Information. Chaos''
implemented as a part of the Basic Research Program at 
National Research University Higher School 
of Economics (HSE University).

\end{document}